\documentclass[a4paper,11pt]{article}
\pdfoutput=1 
\usepackage{jcappub}
\usepackage{amsmath,amssymb}
\usepackage{bm}
\usepackage{xcolor}
\usepackage{graphicx}
\usepackage{enumitem}
\usepackage{geometry}
\usepackage{xspace}
\usepackage{pdflscape}
\usepackage{placeins}
\usepackage{epstopdf}
\usepackage{comment}
\usepackage{subfig}
\makeatletter
\gdef\@fpheader{}
\g@addto@macro\bfseries{\boldmath}
\makeatother

\newcommand{\OmegaGW}{\Omega_{\mathrm{GW}}}
\newcommand{\rhoGW}{\rho_{\mathrm{GW}}}

\let\oldsqrt\sqrt
\def\sqrt{\mathpalette\DHLhksqrt}
\def\DHLhksqrt#1#2{%
\setbox0=\hbox{$#1\oldsqrt{#2\,}$}\dimen0=\ht0
\advance\dimen0-0.2\ht0
\setbox2=\hbox{\vrule height\ht0 depth -\dimen0}%
{\box0\lower0.4pt\box2}}

\newcommand{\dd}{\mathrm{d}}

\newcommand{\sss}[1]{{\scriptscriptstyle{#1}}}
\newcommand{\boldmathsymbol}[1]{{\ensuremath{\boldsymbol{#1}}}}

\newcommand{\uPl}{\mathrm{Pl}}

\newcommand{\ueff}{\mathrm{eff}}

\newcommand{\usssPl}{\sss{\uPl}}

\newcommand{\ud}{\mathrm{d}}

\newcommand{\calH}{\mathcal{H}}

\newcommand{\Mp}{M_\usssPl}

\newcommand{\beq}{\begin{equation}}
\newcommand{\eeq}{\end{equation}}
\newcommand{\bea}{\begin{equation}\begin{aligned}}
\newcommand{\eea}{\end{aligned}\end{equation}}

\newlength{\wsingfig}
\newlength{\wdblefig}
\newlength{\wquadfig}
\newlength{\wtriplefig}
\newcommand{\Eq}[1]{Eq.~(\ref{#1})}

\newcommand{\Fig}[1]{Fig.~{\ref{#1}}}

\newcommand{\Sec}[1]{Sec.~\ref{#1}}

\newcommand{\Hc}[1]{\mathcal{H}}

\renewcommand{\Hc}{\mathcal{H}}
\def\br{\bar\rho}
\def\bp{\bar p}
\def\del{\delta}
\def\ups{\upsilon}
\def\doi{http://doi.org}

\date{today}

\title{No constraints for $f(T)$ gravity from  gravitational waves induced 
from primordial black hole fluctuations}

\author[a]{Theodoros Papanikolaou}

\author[a,b]{Charalampos Tzerefos}

 \author[a,c,d]{Spyros Basilakos}

\author[a,e,f]{Emmanuel N. Saridakis}

\affiliation[a]{National Observatory of Athens, Lofos Nymfon, 11852 Athens, 
Greece}

\affiliation[b]{Department of Physics, National \& Kapodistrian University of 
Athens, Zografou Campus GR 157 73, Athens, Greece}

\affiliation[c]{ Academy of Athens, Research Center for Astronomy and Applied 
Mathematics, Soranou Efesiou 4, 11527, Athens, Greece}

 \affiliation[d]{School of Sciences, European University Cyprus, Diogenes 
Street, Engomi, 1516 Nicosia, Cyprus}
 
\affiliation[e]{CAS Key Laboratory for Researches in Galaxies and Cosmology, 
Department of Astronomy, University of Science and Technology of China, Hefei, 
Anhui 230026, P.R. China}

 \affiliation[f]{School of Astronomy, School of Physical Sciences,
University of Science and Technology of China, Hefei 230026, P.R. China}

\emailAdd{papaniko@noa.gr}
\emailAdd{chtzeref@phys.uoa.gr}
\emailAdd{svasil@academyofathens.gr}
\emailAdd{msaridak@noa.gr}

\abstract{ 
Primordial black hole (PBH) fluctuations can induce a stochastic gravitational wave 
background  at second order, and since this procedure is sensitive to the 
underlying gravitational theory  it  can   be used as a novel tool  to test 
general relativity and extract constraints on possible modified gravity 
deviations. 
We apply this formalism in the framework of $f(T)$ gravity, considering  
three viable mono-parametric models. In particular, we investigate the 
induced modifications   at the level of the gravitational-wave source, which is 
encoded   in terms of the power spectrum of the PBH gravitational potential, as 
well as at the level of their propagation, described in terms of the Green 
function  which  quantifies the propagator of the tensor perturbations.
We find that, within the observationally allowed range of 
  the $f(T)$ model-parameters, the obtained deviations from  general relativity,
both at the levels of source and propagation, are 
practically negligible.  Hence, we conclude that realistic and viable $f(T)$ 
theories  can safely pass the primordial black hole constraints, which may 
offer an 
additional argument in their favor.

}


\begin{document}
\maketitle

\section{Introduction}

Modified gravity is one of the two main avenues that one can follow in order to 
describe the early and late phases of Universe's acceleration 
\cite{CANTATA:2021ktz,Capozziello:2011et}, and compared to the other 
alternative, namely the introduction of the inflaton/dark-energy concept 
\cite{Copeland:2006wr,Cai:2009zp,Martin:2013tda},
has the additional advantage of being closer to  the quantum description of 
gravity \cite{Addazi:2021xuf}. Although the simplest way to build novel classes 
of gravitational theories is to start from the standard, curvature, formulation 
of gravity and extend it in various ways    
\cite{Capozziello:2002rd, Nojiri:2010wj,Nojiri:2005jg,Nicolis:2008in, 
Clifton:2011jh}, one can equally well follow the alternative, torsional, 
formulation   and extend it suitably. In particular, 
since the basic torsional 
theory, namely the   teleparallel equivalent 
of general relativity (TEGR) \cite{Aldrovandi:2013wha, 
Krssak:2018ywd} uses the torsion scalar $T$ as the Lagrangian, one can 
construct torsional modifications extending it such as  in
 $f(T)$ gravity 
\cite{Cai:2015emx,Bengochea:2008gz,Linder:2010py,
Chen:2010va,
 Zheng:2010am,Bamba:2010wb,Cai:2011tc,
Capozziello:2011hj,  
Otalora:2013dsa,Bamba:2013jqa,
Li:2013xea,Ong:2013qja,  
Bamba:2016gbu,Malekjani:2016mtm,Farrugia:2016qqe,
Bahamonde:2017wwk,Karpathopoulos:2017arc,Abedi:2018lkr,DAgostino:2018ngy,
Krssak:2018ywd,
Iosifidis:2018zwo,
Chakrabarti:2019bed, DavoodSadatian:2019pvq,Yan:2019gbw,Wang:2020zfv, 
Bose:2020xdz, Ren:2021tfi,Escamilla-Rivera:2021xql},   in 
$f(T,T_G)$ gravity 
\cite{Kofinas:2014owa,Kofinas:2014daa}, in $f(T,B)$ gravity 
\cite{Bahamonde:2015zma}, in scalar-torsion theories 
\cite{Geng:2011aj,Hohmann:2018rwf}, etc.

 On the other hand, primordial black holes (PBHs), firstly introduced in the 
early 
'70s~\cite{1967SvA....10..602Z, Carr:1974nx,1975ApJ...201....1C}, have 
rekindled 
the interest of the scientific community given the fact that they can solve a 
number of fundamental issues of modern cosmology. In particular, they may 
indeed 
constitute a viable candidate for dark 
matter~\cite{Chapline:1975ojl,Clesse:2017bsw}, and explain the large-scale 
structure formation process through the Poisson fluctuations they can 
seed~\cite{Meszaros:1975ef,Afshordi:2003zb}. At the same time, depending on 
their mass they can give access to a wide variety of physical phenomena from 
the 
early universe up to late times~\cite{Carr:2020gox}.

 PBHs are tightly connected with gravitational wave (GW) physics, and 
specifically 
through the GW signals they are associated with~\cite{Sasaki:2018dmp}. 
In particular, PBHs are connected with GW background signals from PBH merging 
events~\cite{Nakamura:1997sm, Ioka:1998nz, 
Eroshenko:2016hmn, Raidal:2017mfl, Zagorac:2019ekv,Hooper:2020evu} and from PBH 
Hawking radiated gravitons~\cite{Anantua:2008am,Dong:2015yjs}, as 
well as with 
scalar induced gravitational waves, which are induced at second order in 
cosmological perturbation theory either from primordial curvature 
perturbations~\cite{Bugaev:2009zh, Saito_2009, Nakama_2015,Pi:2017gih, 
Yuan:2019udt,Zhou:2020kkf,Fumagalli:2020nvq} (for a recent 
review see \cite{Domenech:2021ztg}) or from Poisson PBH energy density 
fluctuations~\cite{Papanikolaou:2020qtd,Domenech:2020ssp,Kozaczuk:2021wcl}. All 
these signals have been mainly studied within the context of general 
relativity. 

Hence, given  the motivation behind modified gravity theories, one can 
use the aforementioned GW 
portal associated to PBHs in order to constrain them. Specifically,
there have been some first attempts in 
~\cite{Chen:2021nio,Lin:2021vwc}, 
where the authors study the primordial scalar 
induced GWs within the context of Hordenski gravity and non-canonical Higgs 
inflation, as well as in ~\cite{Papanikolaou:2021uhe}
where the scalar induced 
GWs from PBH Poisson fluctuations were studied within $f(R)$ gravity theories 
and in particular within Starobinsky inflation.

In this manuscript, we focus on   $f(T)$ modified gravity, 
  and we study the scalar induced GWs from Poisson fluctuations of 
ultralight PBHs (i.e. with $m_\mathrm{PBH}<10^{9}\mathrm{g}$),
which evaporate before BBN 
and transiently dominate the energy content of the universe before their 
evaporation~\cite{GarciaBellido:1996qt, Hidalgo:2011fj, Martin:2019nuw, 
Zagorac:2019ekv}.
Thus, the main goal of the work is to examine whether 
such an analysis will impose constraints on the various specific $f(T)$ models, 
similarly to other observational investigations 
\cite{Wu:2010mn,Cardone:2012xq,Nesseris:2013jea, 
 Nunes:2016plz,Basilakos:2018arq,
Xu:2018npu,Ren:2022aeo,Huang:2022slc,Zhao:2022gxl}.

The plan of the work is as 
follows: In \Sec{GRPBH}  we review the calculation of the PBH gravitational 
potential of Poisson distributed PBHs within 
general relativity, and in \Sec{teleparallel} we perform the extended
analysis, extracting the PBH gravitational potential in the framework of $f(T)$ 
gravity. In \Sec{SIGW}  we focus on three mono-parametric $f(T)$ gravity models 
and we present the formalism for the computation of the relevant scalar induced gravitational wave signal. Then, in 
\Sec{ConstrfT} we investigate the modifications of the GW signal within $f(T)$ 
gravity by accounting for GW source and GW propagation effects, showing that 
$f(T)$ theories of gravity safely pass the constraints imposed by the 
gravitational-wave signal induced from PBH Poisson fluctuations. Finally, 
\Sec{Conclusions} is devoted to conclusions.

\section{The primordial black hole gravitational potential in general 
relativity}\label{GRPBH}

In this section, we briefly review the calculation of the PBH gravitational 
potential in the case of general relativity, following 
\cite{Papanikolaou:2020qtd,Papanikolaou:2021uhe}. We first present the 
background and perturbation equations, and then we extract 
the  power spectrum of the PBH gravitational potential.

\subsection{Background evolution and scalar perturbations}

We consider a flat 
 Friedmann-Lema\^itre-Robertson-Walker (FLRW) background geometry with  
metric  
\begin{equation}
\label{FLRWmetric_background}
\mathrm{d}s^{2}_\mathrm{b}=-\mathrm{d}t^{2}+a^{2}(t)\delta_{ij}\mathrm{d}x^{i}\mathrm{d}x^{j}\,,
\end{equation}
where $a(t)$ is the scale factor. Additionally,  we consider that the Universe 
is 
filled with hydrodynamic fluid matter described with
energy-momentum tensor $T^{\mathrm{m}}_{\mu\nu} = \mathrm{diag}( - \br , \bp , 
\bp , \bp)$, where $\br$ and $\bp$ are the total matter (i.e. including 
radiation, baryonic and dark matter) energy density and pressure. Thus, the   
Friedmann equations are
    \begin{align}
    H^2&= \frac {8\pi G  }{3} \br + \frac{\Lambda}{3} \equiv \frac {8\pi G  
}{3} \br_{\mathrm{tot}} \label{F1}\\
\dot{H} + H^2 &=-\frac{4\pi G  }{3}  \left(\br + 3\bp \right) + 
\frac{\Lambda}{3} \equiv -\frac{4\pi G  }{3}  \left(\br_{\mathrm{tot}} + 
3\bp_{\mathrm{tot}}\right), \label{F2}
    \end{align}
    where $G$ is  the   Newton constant (throughout this 
paper we work in units where $c=1$),   $\Lambda$  is the 
cosmological constant,   
  $H=\dot{a}/a$ is the Hubble parameter and dots denote derivatives 
with respect to the cosmic time $t$. Note that   we have 
introduced the   total background 
energy density and pressure,
$\br_{\mathrm{tot}}$ and $\bp_{\mathrm{tot}}$, which include   the   total 
matter sector alongside with the cosmological constant. 

In order to proceed to the investigation of the scalar perturbations it proves convenient 
to  introduce the conformal time  $\eta$, defined 
as $\mathrm{d}t \equiv a \mathrm{d}\eta$, and hence the conformal Hubble parameter reads as
$\Hc \equiv a'/a=aH$,
where primes denote derivatives with respect to  $\eta$.  
 Restricting ourselves to scalar 
perturbations in the Newtonian gauge, we can write the   perturbed   
metric    as 
\bea
\label{perturbed FLRW metric with scalar perturbations}
\mathrm{d}s^2 = a^2(\eta)\left\lbrace-(1+2\Psi)\mathrm{d}\eta^2  + 
\left[(1-2\Phi)\delta_{ij}\right]\mathrm{d}x^i\mathrm{d}x^j\right\rbrace \,,
\eea
where
$\Psi$ and $\Phi$ are the two Bardeen potentials  \cite{Bardeen:1980kt}. 
Additionally, we include perturbations around the background stress-energy 
tensor of the total matter content of the Universe (matter and radiation) which 
we express as follows:  
\begin{align}
 T^{0}_{0} &= -(\br + \del \rho)\nonumber \\
 T^{0}_{i} &= (\br + \bp)\ups_{i} \, , \, \, \ups_{i} \equiv a \delta u_{i} 
\nonumber\\
 T^{i}_{j} &= \bp ( \del ^{i}_{j} + \Pi^{i}_{j}), 
\label{hydrodynamicstressenergytensor}
\end{align}
where $\del \equiv \del \rho / \br $ \, is the relative energy density 
perturbation, $\delta u_i \equiv \ups_i/a $ is the velocity perturbation and 
$\Pi^i_j$ is the (dimensionless) anisotropic stress. In this context, one 
obtains the following perturbed equations for $\Phi$ and  $\Psi$ \cite{LL}: 
   \begin{align}
      3\Hc(\Phi' + \Hc\Psi) -\nabla ^2\Phi &= -4\pi G a^2 \, \delta \rho 
\label{E1} \\
        (\Phi' + \Hc\Psi)_{,i} &= 4\pi G a^2 (\br + \bp) \ups_{i} \label{E2} 
\\ 
         \Phi'' + \Hc(\Phi' + 2\Psi') + (\Hc^2 + 2\Hc')\Phi + \nabla ^2(\Phi - 
\Psi) /3   &= 4\pi G a^2  \delta p \label{E3} \\
    \Phi - \Psi  &= 8\pi G a^2 \bp \Pi . \label{E4}
    \end{align}
In the time period that we focus on, 
$\Pi^i_j$ is negligible  and therefore $\Phi \approx \Psi$,
which we consider to be the case from now on. Elaborating on the above 
equations, 
and   using also the  (total) conservation equation  $\bar{\rho} ' = - 
3\Hc (\bar{\rho} + \bar{p})$, one obtains 
 \beq\label{eq:Phi:GR}
\Phi'' + 3\mathcal{H}\left(1+c^2_\mathrm{s}\right)\Phi' - 
c^2_\mathrm{s}\nabla^2\Phi + 3\left(c^2_\mathrm{s} - w \right)\mathcal{H}^2\Phi 
= - \frac{9}{2}c^2_\mathrm{s} (1+ w) \mathcal{H}^3\left( \frac{\delta p}{\bp '} 
- \frac{\delta \rho}{\br '} \right),
\eeq
with  $w \equiv \bp / \br $  being the equation-of-state parameter and $ 
c^2_\mathrm{s}\equiv \bp' / \br' $ the sound speed square of the total matter 
content sector. 
Hence, during the period of PBH domination, $\Phi$ is the 
potential arising from the PBH distribution.

\subsection{The PBH gravitational potential power 
spectrum}\label{sec:PowerSpectrumPhiGR}
In order to proceed we assume conventionally that PBHs are formed in the radiation-dominated era. Considering that PBHs are randomly distributed, their energy density is inhomogeneous while the total background energy density is homogeneous,
and thus their energy density perturbations can be viewed as isocurvature Poisson fluctuations with the associated Poissonian power spectrum 
for the PBH density contrast being read as
\cite{Papanikolaou:2020qtd}
\beq
\label{eq:PowerSpectrum:PBH}
\mathcal{P}_\delta(k) = \frac{k^3}{2\pi^2}P_\delta(k)= \frac{2}{3\pi} 
\left(\frac{k}{k_{\mathrm{UV}}}\right)^3 \Theta(k_\mathrm{UV}-k),
\eeq
where we have assumed
monochromatic 
PBH mass function \cite{MoradinezhadDizgah:2019wjf}.
Additionally,   $k_{\mathrm{UV}}\equiv a/\bar{r}$ is the Ultraviolet (UV) 
cut-off scale, which is
related to the 
mean PBH separation scale, since at smaller scales  the PBH fluid description 
is not valid.  Introducing then the density parameter for the PBH fluid as 
$\Omega_\mathrm{PBH}\equiv 
\frac{\rho_\mathrm{PBH}}{\rho_\mathrm{tot}}$ we can find that  during the 
radiation-dominated era,  $\Omega_\mathrm{PBH} \propto a$, and therefore if the 
initial  PBH   abundance is sufficiently large then PBHs can 
  dominate. Hence, the isocurvature PBH 
  perturbations in the radiation-dominated era will be converted to 
  adiabatic curvature perturbations in the subsequent PBH dominated era  
\cite{Kodama:1986fg,Kodama:1986ud}.

In order to relate   $\Phi$ and  $\delta_\mathrm{PBH}$,  we introduce    the 
uniform-energy density curvature perturbation of each fluid, $\zeta_i$\cite{Wands:2000dp}.
At super-horizon scales,  $\zeta_\mathrm{r}$ and  $\zeta_\mathrm{PBH}$ are 
 separately conserved, hence 
at the PBH formation time we can neglect the adiabatic 
contribution associated to the radiation fluid on the scales we are interested in and obtain
\bea
\label{eq:zeta:delta:superH}
\zeta\simeq \frac{1}{3} \delta_\mathrm{PBH}(t_\mathrm{f})\quad 
\mathrm{for}\quad 
k\ll \mathcal{H}\,. 
\eea
Thus,  during the PBH-matter dominated era, where $w=0$ and 
$\Phi$ is constant in time~\cite{Mukhanov:1990me} using the fact that on 
super-horizon scales $\zeta \simeq - \mathcal{R}$~\cite{Wands:2000dp}, where 
$\mathcal{R}$ is the comoving curvature perturbation, as well as the relation 
between $\mathcal{R}$ and $\Phi$ in GR~\cite{Mukhanov:1990me}, we finally 
obtain 
that 
\cite{Papanikolaou:2020qtd,Papanikolaou:2021uhe}
\bea
\label{eq:Phi:delta:superH00}
\Phi\simeq -\frac{1}{5} \delta_\mathrm{PBH}(t_\mathrm{f})\quad 
\mathrm{for}\quad 
k\ll \mathcal{H}\,.
\eea

On the other hand, at sub-horizon scales one can determine the evolution of 
$\delta_
\mathrm{PBH}$ by solving the evolution equation for the matter density 
perturbations 
\begin{equation}
 \label{eq:growth:equation:GR}
\delta^{\prime\prime}_\mathrm{m}+\calH\delta^\prime_\mathrm{m}-4\pi 
G 
a^2\bar{\rho}_\mathrm{m}\delta_\mathrm{m} = 0 ,
\end{equation}
which, in the case of a Universe with radiation and PBH-matter, takes the form 
of the so-called M\'eszaros growth equation~\cite{Meszaros:1974tb}:
\bea\label{eq:Meszaros in GR}
\frac{\dd^2 \delta_\mathrm{PBH}}{\dd s^2}+\frac{2+3s}{2s(s+1)}\frac{\dd 
\delta_\mathrm{PBH}}{\dd s}-\frac{3}{2s (s+1)} \delta_\mathrm{PBH}=0\,.
\eea 
By solving  the M\'eszaros growth 
equation 
we deduce that 
the dominant solution deep in the PBH-dominated era is 
~\cite{Papanikolaou:2020qtd,Papanikolaou:2021uhe}
\beq\label{eq:delta_PBH_sub_GR}
\delta_\mathrm{PBH}\simeq \frac{3s}{2}
\delta_\mathrm{PBH}(t_\mathrm{f}).
\eeq
 Since   the Bardeen potential 
is related to  the density contrast through the Poisson equation,   one gets 
that in the 
matter-dominated era
\bea
\label{eq:delta:Phi:MD}
\delta_\mathrm{PBH} =  -\frac{2}{3} \left(\frac{k}{\mathcal{H}}\right)^2\Phi,
\eea
and thus inserting into \Eq{eq:delta_PBH_sub_GR}  we find
\bea
\label{eq:Phi:delta:subH}
\Phi\simeq -\frac{9}{4}\left(\frac{\calH_{\mathrm{d}}}{k}\right)^2\, 
\delta_\mathrm{PBH}(t_\mathrm{f})
\quad \mathrm{for}\quad k\gg {\calH}_{\mathrm{d}}\, ,
\eea
with ${\calH}_{\mathrm{d}}$ being the conformal Hubble parameter at the
PBH domination time.
At the end, interpolating between (\ref{eq:Phi:delta:subH}) and 
(\ref{eq:Phi:delta:superH00}),  and using (\ref{eq:PowerSpectrum:PBH}), we obtain that
\beq\label{eq:PowerSpectrum:Phi:PBHdom}
\mathcal{P}_\Phi(k) \equiv\frac{k^3}{2\pi^2}P_\Phi(k)= 
\frac{2}{3\pi}\left(\frac{k}{k_\mathrm{UV}}\right)^3 
\left(5+\frac{4}{9}\frac{k^2}{k_{\mathrm{d}}^2}\right)^{-2}\, ,
\eeq
with $k_\mathrm{d}\equiv \cal{H}_\mathrm{d}$ being the comoving scale exiting the 
Hubble radius at PBH 
domination time.

\section{The primordial black hole gravitational potential in $f(T)$ 
gravity}
\label{teleparallel}

In this section we perform the calculation of the  PBH gravitational potential in the framework of $f(T)$ gravity.  We first review the relevant  background and perturbation equations and then we proceed to calculate the associated power spectrum.

\subsection{Background evolution and scalar perturbations}

In the torsional formulation of gravity one uses the tetrad or vierbein fields 
${\mathbf{e}_A(x^\mu)}$ as the dynamical 
variables instead of the metric tensor. They form an orthonormal basis for the 
tangent space at each point  $x^\mu$ of the manifold, that is $ e_A \cdot e_B = 
\eta_{AB}$, where  Greek  and Latin indices run in coordinate 
and tangent space respectively and $\eta_{AB}={\rm diag} 
(-1,1,1,1)$  is the Minkowski metric for 
the (flat) tangent space. One can express them in the coordinate basis
$\mathbf{e}_A=e^\mu_A\partial_\mu $ and thus construct the metric as  
\begin{equation}  \label{metricrel}
g_{\mu\nu}(x)=\eta_{AB}\, e^A_\mu (x)\, e^B_\nu (x).
\end{equation}
\noindent 
In teleparallel gravity  one describes gravity using the torsion of spacetime 
instead of its curvature. In this spirit, instead of the familiar 
Christoffel connection, which is the unique connection whose torsion vanishes, 
one can introduce the  
 Weitzenb\"{o}ck connection  
$\overset{\mathbf{w}}{\Gamma}^\lambda_{\nu\mu}\equiv e^\lambda_A\:
\partial_\mu
e^A_\nu$, which is a connection whose curvature vanishes  
\cite{Aldrovandi:2013wha}. The torsion tensor is given by  
\begin{equation}
\label{torsten}
{T}^\lambda_{\:\mu\nu}\equiv\overset{\mathbf{w}}{\Gamma}^\lambda_{
\nu\mu}-%
\overset{\mathbf{w}}{\Gamma}^\lambda_{\mu\nu}
=e^\lambda_A\:(\partial_\mu
e^A_\nu-\partial_\nu e^A_\mu),
\end{equation}
and  its  
contraction provides the torsion scalar  as
\begin{equation}
\label{torsiscal}
T\equiv\frac{1}{4}
T^{\rho \mu \nu}
T_{\rho \mu \nu}
+\frac{1}{2}T^{\rho \mu \nu }T_{\nu \mu\rho }
-T_{\rho \mu }^{\ \ \rho }T_{\
\ \ \nu }^{\nu \mu }.
\end{equation}
Using $T$ as a Lagrangian gives rise to the teleparallel equivalent of general 
relativity (TEGR), since variation in terms of the tetrads leads to the same 
field
equations with general relativity \cite{Aldrovandi:2013wha}.

One then can generalize TEGR by extending $T$ to an arbitrary function of $T$ as the 
Lagrangian, resulting to $f(T)$ gravity, whose action is 
\cite{Cai:2015emx}
\begin{equation} 
S = \frac{1}{16\pi G} \int \mathrm{d}^4x \, |e| \, f(T) + \int \mathrm{d}^4x \, |e| \, 
\mathcal{L}_\mathrm{m} \label{ft},
\end{equation}
with $|e| = 
\mathrm{det}(e^{A}_{\mu}) = \sqrt {-g}$,
  and where we have   
included 
the total matter Lagrangian $L_m$ for completeness.  
 Variation 
of the action (\ref{ft}) with respect to the tetrad $e^{A}_{\mu}$ 
yields the following field equations:
\begin{equation}\label{equationsom}
 e^{-1}\partial_{\mu}(ee_A^{\rho}S_{\rho}{}^{\mu\nu})f_{T}
 +
e_A^{\rho}S_{\rho}{}^{\mu\nu}\partial_{\mu}({T})f_{TT} 
-f_{T}e_{A}^{\lambda}T^{\rho}{}_{\mu\lambda}S_{\rho}{}^{\nu\mu}+\frac{1}{4} 
e_ {A
} ^ {
\nu
}f({T})  
= 4\pi Ge_{A}^{\rho}\overset {\mathbf{em}}T_{\rho}{}^{\nu},
\end{equation}
  with $f_{T}\equiv\partial f/\partial T$,
$f_{TT}\equiv\partial^{2} 
f/\partial T^{2}$
and where $\overset{\mathbf{em}}{T}_{\rho}{}^{\nu}$  denotes  the total 
matter
energy-momentum 
tensor. Note that for convenience  we have introduced the super-potential tensor
$
S_\rho^{\:\:\:\mu\nu}\equiv\frac{1}{2}\Big(K^{\mu\nu}_{\:\:\:\:\rho}
+\delta^\mu_\rho
\:T^{\alpha\nu}_{\:\:\:\:\alpha}-\delta^\nu_\rho\:
T^{\alpha\mu}_{\:\:\:\:\alpha}\Big)$, with 
$K^{\mu\nu}_{\:\:\:\:\rho}\equiv-\frac{1}{2}\Big(T^{\mu\nu}_{
\:\:\:\:\rho}
-T^{\nu\mu}_{\:\:\:\:\rho}-T_{\rho}^{\:\:\:\:\mu\nu}\Big)$  being the  
contorsion 
tensor.

Applying $f(T)$ gravity in a cosmological framework we impose the  FLRW metric 
(\ref{FLRWmetric_background}), which in turn arises from the   tetrad
$e_{\mu}^A={\rm
diag}(1,a,a,a)$. Inserting this ansatz  
into (\ref{equationsom}) one obtains the familiar Friedmann equations
    \begin{align}
    H^2&= \frac {8\pi G  }{3}\left( \br+\rho_{\mathrm{f(T)}}\right)   \equiv 
\frac 
{8\pi G  }{3} 
\br_{\mathrm{tot}} \label{F1}\\
\dot{H} + H^2 &=-\frac{4\pi G  }{3}  \left(\br + 3\bp +\rho_{\mathrm{f(T)}}  
+3p_{\mathrm{f(T)}} \right)  \equiv -\frac{4\pi G  }{3}  
\left(\br_{\mathrm{tot}} + 
3\bp_{\mathrm{tot}}\right), \label{F2}
    \end{align}
    where we have defined the effective energy density and pressure due to the 
$f(T)$ modification as \cite{Cai:2015emx}
    \begin{eqnarray}
&&\rho_{\mathrm{f(T)}}\equiv\frac{3}{8\pi
G}\left[-\frac{f}{6}+\frac{Tf_T}{3}-\frac{T}{6}\right], \label{rhoDDE}\\
\label{pDE}
&&p_{\mathrm{f(T)}}\equiv\frac{1}{16\pi G}\left[\frac{f-f_{T} T
+2T^2f_{TT}}{ f_{T}+2Tf_{TT}}\right],
\end{eqnarray}
with equation-of-state parameter given by 
    \begin{eqnarray}
&&w_{\mathrm{f(T)}}\equiv \frac{p_{\mathrm{f(T)}}}{\rho_{\mathrm{f(T)}}}= 
 \frac{f-f_{T} T
+2T^2f_{TT}}{ (f_{T}+2Tf_{TT})(-f+2  Tf_T- T)} .
\label{wdeft}
\end{eqnarray}
Thus, $\br_{\mathrm{tot}}\equiv \br +  \rho_{\mathrm{f(T)}} $ and  
$\bp_{\mathrm{tot}}\equiv \bp + p_{\mathrm{f(T)}} $.
Lastly, note that in FLRW geometry according to 
(\ref{torsiscal}) the torsion scalar  becomes simply $T=6H^2$.

 Proceeding to the perturbation level and focusing on scalar perturbations, one 
can write the perturbed tetrad fields as follows:
\begin{eqnarray}
 e_{\mu}^A = \bar{e}_{\mu}^A + {\mathbb{E}}_{\mu}^A ~,
 \label{basicveirpertChen}
\end{eqnarray}
where the symbols ${e}_{\mu}^A$ and $\bar{e}_{\mu}^A$ are used for the 
perturbed 
and the unperturbed tetrad fields correspondingly. Then, one can impose the following 
ansatz for the scalar contributions of the perturbed tetrad fields:
\begin{eqnarray}
 \label{Chen_pert1}
 \bar{e}_{\mu}^0 = \delta_{\mu}^0, \,\,\,\,\, \bar{e}_{\mu}^a = \delta_{\mu}^aa,
\,\,\,\,\,
 \bar{e}^{\mu}_0 = \delta^{\mu}_0, \,\,\,\,\, \bar{e}^{\mu}_a = 
\frac{\delta^{\mu}_a}{a}
~,
\end{eqnarray}
and
\begin{eqnarray}
 \label{Chen_pert2}
 \! \! \!\!{\mathbb{E}}_{\mu}^0 = \delta_{\mu}^0\Psi, \,\,\, 
{\mathbb{E}}_{\mu}^a
=-\delta_{\mu}^a a\Phi, \,
\,\,
 {\mathbb{E}}^{\mu}_0 = -\delta_{0}^{\mu}\Psi, \,\,\, {\mathbb{E}}^{\mu}_a =
\frac{\delta^{\mu}_
a}{a}\Phi .\
\end{eqnarray}
In the above expressions, the scalar perturbations $\Phi$ and
$\Psi$ are introduced, which are functions of space ${\bf x}$ and time $t$. 
With such a choice, one can match the tetrad perturbations with a perturbed metric in the Newtonian gauge 
\cite{Chen:2010va,Bahamonde:2021gfp},
 namely
\begin{eqnarray}\label{pertmetric}
 \mathrm{d}s^2 = -(1 + 2\Psi) \mathrm{d}t^2 +a^2(1-2\Phi)\delta_{ij} 
\mathrm{d}x^i \mathrm{d}x^j.
\end{eqnarray}

Following then \cite{Cai:2015emx}, and writing $f(T)= T+ F(T)$, 
one
can expand the gravitational equations of motion (\ref{equationsom}) to linear 
order and obtain the $(00)$, 
$(0i)$, $(
ij)$ and $(ii)$ component perturbation equations, which read as follows:
\begin{equation}
\label{delta_00_fT_bounce}
 \left( 1+F_{T} \right) \frac{\nabla^2}{a^2}\Psi -3(1+F_{T})H\dot\Psi
 -3(1+F_{T})H^2\Phi 
 +36F_{TT}H^3(\dot\Psi+H\Phi) = 4{\pi}G ~ \delta\rho ~,
\end{equation}
\begin{align}
\label{delta_0i_fT_bounce}
 (1+F_{T}-12H^2F_{TT}) (\dot\Psi+H\Phi) = 4{\pi}G (\bar{\rho} + 
\bar{p})\upsilon 
~,
\end{align}
\begin{align}
\label{delta_ij_fT_bounce}
 (1+F_{T})(\Psi-\Phi) = 8{\pi}G ~\bar{p}\Pi~,
\end{align}
and
\begin{align}
\label{delta_ii_fT_bounce}
 & (1+F_{T}-12H^2F_{TT})\ddot\Psi +H(1+F_{T}-12H^2F_{TT})\dot\Phi  \nonumber\\
 & +3H(1+F_{T} -12H^2F_{TT} -12\dot{H}F_{TT}+48H^2\dot{H}F_{TTT})\dot\Psi 
\nonumber\\
 & +\left[3H^2(1+F_{T}-12H^2F_{TT}) 
 +2\dot{H}(1+F_{T}-30H^2F_{TT} +72H^4F_{TTT})\right]\Phi
\nonumber\\
 & +\frac{1+F_{T}}{3a^2}\nabla^2(\Psi-\Phi) = 4{\pi}G ~\delta{p} ~,
\end{align}
where the total matter content of the Universe is expressed as in 
(\ref{hydrodynamicstressenergytensor}).

In the context of this work, all the models of $f(T)$ gravity that we are going 
to investigate (see \Sec{ConstrfT}) are the viable ones, characterised by $F_T\ll 1$, $F_{TT}\ll 
1$ and $F_{TTT}\ll 1$ \cite{Nesseris:2013jea}. 
Furthermore, as 
we already mentioned in the previous section, the anisotropic stress can be 
neglected, and therefore by virtue of (\ref{delta_ij_fT_bounce}) one obtains 
that $\Phi \sim 
\Psi$. Therefore, under these conditions it is straightforward to show that 
from 
the form of the aforementioned equations one can use for the evolution of 
$\Phi$ 
the GR equation (\ref{eq:Phi:GR}), namely:
\beq\label{eq:Phi:ft}
\Phi'' + 3\mathcal{H}\left(1+c^2_\mathrm{s}\right)\Phi' - 
c^2_\mathrm{s}\nabla^2\Phi + 3\left(c^2_\mathrm{s} - w \right)\mathcal{H}^2\Phi 
= - \frac{9}{2}c^2_\mathrm{s} (1+ w) \mathcal{H}^3\left( \frac{\delta p}{\bp '} 
- \frac{\delta \rho}{\br '} \right).
\eeq
On the other hand, we stress out that the effect of $f(T)$ gravity will be taken 
into account at the level of the evolution of the matter density contrast, as 
we will show in the next subsection, as well as at the level of the tensor 
perturbations (see \Sec{SIGW}).   

Regarding  the evolution of the matter density contrast, it is given via the 
growth equation, which is derived from (\ref{delta_00_fT_bounce}) and 
(\ref{delta_0i_fT_bounce}) at  subhorizon scales, and assuming matter 
domination  it is given by 
\cite{Nesseris:2013jea,Anagnostopoulos:2019miu}
\beq \label{gfT}
\delta^{\prime\prime}_\mathrm{m}+\calH\delta^\prime_\mathrm{m}-4\pi 
G_\mathrm{eff} 
a^2\bar{\rho}_\mathrm{m}\delta_\mathrm{m} = 0.
\eeq
In this expression the quantity
  \begin{eqnarray}
\label{Geff}
 G_{\rm eff}\equiv \frac{{G}}{f_{T}} 
\end{eqnarray}
is the effective Newton's constant and primes denote derivatives with respect 
to 
the conformal time. Thus, comparing (\ref{eq:growth:equation:GR}) and 
(\ref{gfT}), one can see that the effect of $f(T)$ gravity at the level of the
matter perturbations at sub-horizon scales is essentially captured by the 
modification of the Newton's constant, which is related to the modification of the 
gravitational field.  As we can see,  in the limit 
$F(T)\rightarrow const.=-2\Lambda$ all the above equations recover the ones of 
$\Lambda$CDM cosmology.

\subsection{The Power Spectrum of the PBH Gravitational Potential in $f(T)$ 
gravity}

 We shall now repeat the procedure of subsection 
\ref{sec:PowerSpectrumPhiGR} 
but
in the context of $f(T)$ gravity. Once again, we will use  the 
uniform-energy density curvature perturbation of each fluid, $\zeta_i$, in 
order 
to relate $\Phi$ and  $\delta_\mathrm{PBH}$. For $\zeta$ we will use the usual 
definition \cite{Wands:2000dp}:

\begin{equation}
    \zeta \equiv -\Phi - \Hc \frac{\delta \rho}{\bar{\rho}'} \label{zeta}.
\end{equation}
The relevant fluids for our analysis are the radiation and the PBH-matter ones. 
Their 
corresponding energy-momentum tensors are both conserved, i.e. the (background) 
continuity 
equation  $\bar{\rho}_{i} ' = - 3\Hc (\bar{\rho}_{i} + \bar{p}_{i})$ holds for 
each fluid, and thus $\zeta_{i}$ is expressed as:
\begin{equation}
  \zeta_{i} \equiv -\Phi + \frac{\delta_{i}}{3(1+w_{i})}. \label{z}  
\end{equation}
By substituting the equation of state   $w_{i}$ for each fluid 
we obtain:
\beq\label{eq:zeta_r}
\zeta_\mathrm{r}=-\Phi+\frac{1}{4} \delta_\mathrm{r},
\eeq
\beq\label{eq:zeta_PBH}
\zeta_\mathrm{PBH}=-\Phi+\frac{1}{3} \delta_\mathrm{PBH}.
\eeq
Moreover, we introduce the isocurvature perturbation defined as:
\beq\label{eq:S definition}
S = 3\left(\zeta_\mathrm{PBH}-\zeta_\mathrm{r}\right) =\delta_\mathrm{PBH} - 
\frac{3}{4} \delta_\mathrm{r} \,.
\eeq

On superhorizon scales,  $\zeta_\mathrm{r}$ and  $\zeta_\mathrm{PBH}$ are 
conserved separately  ~\cite{Wands:2000dp},  like the isocurvature perturbation 
$S$. Thus, in the PBH-dominated era, $\zeta\simeq \zeta_\mathrm{PBH} = 
\zeta_{\mathrm{r}}+S/3 \simeq S/3$. Since $S$ is conserved, it can be 
calculated 
at formation time $t_\mathrm{f}$. Consequently, neglecting  the adiabatic 
contribution associated to the radiation fluid at the PBH formation time, since 
it does not play any role at the scales considered here,  from 
\Eq{eq:S definition} we obtain that $S=\delta_\mathrm{PBH}(t_\mathrm{f})$. 
Hence, as in the case of GR   we find  that
\bea
\label{eq:zeta:delta:superH}
\zeta\simeq \frac{1}{3} \delta_\mathrm{PBH}(t_\mathrm{f})\quad \mathrm{if}\quad 
k\ll \mathcal{H}\,. 
\eea

Further, as we show explicitly in Appendix \ref{AppendixB}, 
  in the context of $f(T)$ gravity the  property  $\zeta\simeq -\mathcal{R}$ 
is valid too at super-horizon scales as it does in GR (see e.g. 
\cite{Wands:2000dp}),  by requiring that $F_T\ll 1$, with $\mathcal{R}$ being the 
comoving curvature 
perturbation defined in the usual way
\bea
\label{eq:zeta:Bardeen}
\mathcal{R}  \equiv
\frac{2}{3}\frac{{\Phi}^\prime/\mathcal{H}+\Phi}{1+w }+\Phi\, .
\eea
During a matter-dominated 
era, such as the one driven by PBHs, $\Phi^\prime $ can be neglected since it 
is 
proportional to the decaying mode, thus we obtain  
$\mathcal{R}=-\zeta=(5/3)\Phi$. Therefore, combining with 
(\ref{eq:zeta:delta:superH}), we deduce that
\bea
\label{eq:Phi:delta:superH}
\Phi\simeq -\frac{1}{5} \delta_\mathrm{PBH}(t_\mathrm{f})\quad \mathrm{if}\quad 
k\ll \mathcal{H}\,.
\eea

Let us now focus on sub-Hubble scales. We can determine the evolution of 
$\delta_
\mathrm{PBH}$ by solving the evolution equation of the matter density 
perturbations in $f(T)$ gravity (\ref{gfT}).
At the background level, the Friedmann equation (\ref{F1}) can be expressed 
as 
$
\calH^2 = \frac{8\pi 
Ga^2}{3}\left[\bar{\rho}_\mathrm{PBH}+\bar{\rho}_\mathrm{r}+ 
\rho_\mathrm{f(T)}\right]$
where $\br^\mathrm{f(T)}$ is given by (\ref{rhoDDE}). 
Since at the epochs we focus on, namely before Big Bang Nucleosynthesis 
(BBN), we expect that deviations from $\Lambda$CDM are negligible, we can   
neglect  the 
effective fluid contribution, writing  the Friedmann 
equation as 
\beq\label{eq:Friedmann:Meszaros:DE_neglected}
\calH^2 \simeq 
\calH^2_\mathrm{f}\Omega^2_\mathrm{PBH,f}\left(\frac{1}{s}+\frac{1}{s^2}\right).
\eeq
In this equation,  $s\equiv a/a_\mathrm{d}$ and $a_\mathrm{d}$ denotes the 
time at the 
transition from the radiation to the PBH domination era, while we have 
assumed that $\Omega_\mathrm{r,f}=1$ since PBHs are considered to be formed in 
the radiation era \cite{Papanikolaou:2020qtd}. Note that the scale factor is 
normalised at one at formation time, i.e. $a_\mathrm{f}=1$.  

At the perturbation level, we can treat 
the gas of PBHs as a matter fluid, and by using $s$ as  the time variable, the 
growth 
equation (\ref{gfT}) can be recast in the following form:
\bea\label{eq:Meszaros in f(T):subhorizon}
\frac{\dd^2 \delta_\mathrm{PBH}}{\dd s^2}+\frac{2+3s}{2s(s+1)}\frac{\dd 
\delta_\mathrm{PBH}}{\dd s}-\frac{3}{2s (s+1)} \frac{1}{
f_T}  
\delta_\mathrm{PBH}=0\,.
\eea
  We proceed by relating  our solution for $\delta_\mathrm{PBH}$ from 
(\ref{eq:Meszaros in f(T):subhorizon}) with $\Phi$, via the sub-Hubble scale 
approximation of the time-time field equation in $f(T)$ gravity for the PBH 
dominated era (equation (\ref{delta_00_fT_bounce})), which is:
\bea\label{eq:Phi:delta:subH:f(T)}
\delta_\mathrm{PBH} = -\frac{2}{3} \left( \frac{k}{\mathcal{H}}\right)^2 
f_T\Phi.
\eea
Hence, making   an interpolation between \Eq{eq:Phi:delta:superH} and  
\Eq{eq:Phi:delta:subH:f(T)}, as in the case of GR, and using the expression for 
the PBH matter power spectrum in \Eq{eq:PowerSpectrum:PBH}, we straightforwardly 
extract the following PBH gravitational potential power spectrum:
\beq\label{eq:PowerSpectrum:Phi:PBHdom:f(T)}
\mathcal{P}_\Phi(k) \equiv\frac{k^3}{2\pi^2}P_\Phi(k) = 
\frac{2}{3\pi}\left(\frac{k}{k_\mathrm{UV}}\right)^3\left[5+\frac{2}{3}
\left(\frac{k}
{\mathcal{H}}\right)^2 \frac{f_T}{\xi(a)}\right]^{-2}.
\eeq
In the above expression, $\xi(a)$ is defined as
\beq\label{eq:xi_definition}
\xi(a)\equiv \frac{\delta_\mathrm{PBH}(a)}{\delta_\mathrm{PBH}(a_\mathrm{f})},
\eeq 
where $\delta_\mathrm{PBH}(a)$ is the solution of
\Eq{eq:Meszaros in f(T):subhorizon}. As verified numerically, $\xi(a)$ has a 
mild 
dependence on the comoving scale $k$, and thus for practical reasons we will 
consider $\xi(a)$ as $k$ independent.


\section{Scalar induced gravitational waves in $f(T)$ gravity  }
\label{SIGW}

Since we have calculated    the power spectrum of the gravitational 
potential of the initially Poisson-distributed PBHs, we can now proceed to the 
extraction of the  stochastic gravitational wave background induced  from the 
PBH Poisson fluctuations.

\subsection{Tensor Perturbations}
\label{subsec:tensor_perturbations}

The perturbed metric in the Newtonian gauge, assuming as mentioned above zero 
anisotropic stress, can be recast as
\bea
\label{metric decomposition with tensor perturbations}
\mathrm{d}s^2 = a^2(\eta)\left\lbrace-(1+2\Phi)\mathrm{d}\eta^2  + 
\left[(1-2\Phi)\delta_{ij} + 
\frac{h_{ij}}{2}\right]\mathrm{d}x^i\mathrm{d}x^j\right\rbrace,
\eea
where we have multiplied by a factor $1/2$ the second-order tensor perturbation 
as it is standard in the literature. Then, by 
Fourier transforming the tensor perturbations and taking into account the two
polarization modes of the GWs in $f(T)$ gravity \cite{Bamba:2013ooa}, namely 
the 
$\times$ and the 
$+$ 
as in GR case, the equation of 
motion for the tensor modes $h_\boldmathsymbol{k}$ reads as
\beq
\label{Tensor Eq. of Motion}
h_\boldmathsymbol{k}^{s,\prime\prime} + 
2\mathcal{H} (1-\gamma_T)  h_\boldmathsymbol{k}^{s,\prime} + k^{2}  
h^s_\boldmathsymbol{k} = 4 S^s_\boldmathsymbol{k}\, ,
\eeq
with   $s = (+), (\times)$. In this equation the modified dispersion due to 
the $f(T)$ effects is quantified by the term \cite{Cai:2018rzd}
\beq
\label{Tensor Eq. of Motion}
 \gamma_T \equiv -\frac{f_T'}{2\mathcal{H} f_T},
\eeq
while the source function $S^s_\boldmathsymbol{k}$ is given by
\beq
\label{eq:Source:def}
S^s_\boldmathsymbol{k}  = \int\frac{\mathrm{d}^3 
\boldmathsymbol{q}}{(2\pi)^{3/2}}e^s_{ij}(\boldmathsymbol{k})q_iq_j\left[
2\Phi_\boldmathsymbol{q}\Phi_\boldmathsymbol{k-q} + 
\frac{4}{3(1+w_\mathrm{tot})}(\mathcal{H}^{-1}\Phi_\boldmathsymbol{q} 
^{\prime}+\Phi_\boldmathsymbol{q})(\mathcal{H}^{-1}\Phi_\boldmathsymbol{k-q} 
^{\prime}+\Phi_\boldmathsymbol{k-q}) \right],
\eeq
where the polarization tensors  
$e^{s}_{ij}(k)$ are defined as \cite{Capozziello:2011et}
\beq
e^{(+)}_{ij}(\boldmathsymbol{k}) = \frac{1}{\sqrt{2}}
\begin{pmatrix}
1 & 0 & 0\\
0 & -1 & 0 \\ 
0 & 0 & 0 
\end{pmatrix}, \quad
e^{(\times)}_{ij}(\boldmathsymbol{k}) = \frac{1}{\sqrt{2}}
\begin{pmatrix}
0 & 1 & 0\\
1 & 0 & 0 \\ 
0 & 0 & 0 
\end{pmatrix}. 
\eeq 
As we have mentioned above,  since we focus on second-order effects, in this 
work we assume that the background evolution is close to that of the
$\Lambda\mathrm{CDM}$ scenario - note that we consider PBH domination eras 
before BBN time. Considering also the fact that in the time period that we 
investigate
the Universe is matter (i.e. PBH) dominated, we have    
$c^2_\mathrm{tot} \approx w_\mathrm{tot} \simeq w_\mathrm{PBH}= 0$.
Hence, for the time evolution of the potential $\Phi$ given by \Eq{eq:Phi:ft},  
we 
obtain
\bea
\label{Bardeen potential 2}
\Phi_\boldmathsymbol{k}^{\prime\prime} + 
\frac{6(1+w_\mathrm{tot})}{1+3w_\mathrm{tot}}\frac{1}{\eta}\Phi_\boldmathsymbol{
k}^{\prime} + w_\mathrm{tot}k^2\Phi_\boldmathsymbol{k} =0\, .
\eea
The solution of the above equation is a superposition of a constant and a 
decaying mode. 
In the late-time limit, one can neglect the 
decaying mode, and write the solution for the Fourier transform of $\Phi$ as 
$\Phi_\boldmathsymbol{k}(\eta) = T_\Phi(\eta) \phi_\boldmathsymbol{k}$, where 
$\phi_\boldmathsymbol{k}$ is the value of the gravitational potential at some 
reference time (which here we consider   to be the time at which PBHs dominate 
the energy budget of the Universe,  $x_\ud$) and $T_\Phi(\eta)$ is a transfer 
function, defined as the ratio of the dominant mode between the times $x$ and 
$x_\ud$.  Consequently, \Eq{eq:Source:def} can be written in a more compact 
form 
as
\beq
\label{Source}
S^s_\boldmathsymbol{k}  =
\int\frac{\mathrm{d}^3 
q}{(2\pi)^{3/2}}e^{s}(\boldmathsymbol{k},\boldmathsymbol{q})F(\boldmathsymbol{q}
,\boldmathsymbol{k-q},\eta)\phi_\boldmathsymbol{q}\phi_\boldmathsymbol{k-q}\, ,
\eeq
where
\bea
\label{F}
\!\!\!\!\!
F(\boldmathsymbol{q},\boldmathsymbol{k-q},\eta) & \equiv 
2T_\Phi(q\eta)T_\Phi\left(|\boldmathsymbol{k}-\boldmathsymbol{q}|\eta\right)  + 
\frac{4}{3(1+w)}\left[\mathcal{H}^{-1}qT_\Phi^{\prime}
(q\eta)+T_\Phi(q\eta)\right]
\\  & \kern-2em
\ \ \ \ \ \ \ \ \ \ \  \ \ \ \ \ \  \ \ \ \ \ \ \ \ \ \ \ \ \ \ \ \ \ \ \ \ 
\cdot \left[\mathcal{H}^{-1}\vert\boldmathsymbol{k}-\boldmathsymbol{q}\vert 
T_\Phi^{\prime}\left(|\boldmathsymbol{k}-\boldmathsymbol{q}
|\eta\right)+T_\Phi\left(|\boldmathsymbol{k}-\boldmathsymbol{q}
|\eta\right)\right],
\eea
and the contraction  $e^s_{ij}(\boldmathsymbol{k})q_iq_j \equiv 
e^s(\boldmathsymbol{k},\boldmathsymbol{q})$ can be expressed in terms of the 
spherical coordinates $(q,\theta,\varphi)$ of the vector $\bm{q}$ as 
\beq
e^s(\boldmathsymbol{k},\boldmathsymbol{q})=
\begin{cases}
\frac{1}{\sqrt{2}}q^2\sin^2\theta\cos 2\varphi \mathrm{\;for\;} s= (+)\\
\frac{1}{\sqrt{2}}q^2\sin^2\theta\sin 2\varphi  \mathrm{\;for\;} s= (\times)  
\end{cases}
\, .
\eeq
Finally,  the solution of  \Eq{Tensor Eq. of Motion} for the tensor modes 
$h^s_\boldmathsymbol{k}$ can be obtained using the Green's function formalism 
where one can write for $h^s_\boldmathsymbol{k}$ that
\bea
\label{tensor mode function}
a(\eta)h^s_\boldmathsymbol{k} (\eta)  =4 
\int^{\eta}_{\eta_\mathrm{d}}\mathrm{d}\bar{\eta}\,  
G^s_\boldmathsymbol{k}(\eta,\bar{\eta})a(\bar{\eta})S^s_\boldmathsymbol{k}(\bar{
\eta}),
\eea
and
where the Green's function  $G^s_{\bm{k}}(\eta,\bar{\eta})$ is the solution of 
the homogeneous equation 
\beq
\label{eq:Green function equation in f(T)}
G_\boldmathsymbol{k}^{s,\prime\prime}(\eta,\bar{\eta})  - 2\Hc \gamma_T 
G_\boldmathsymbol{k}^{s,\prime}(\eta,\bar{\eta}) + \left( k^{2}  
-\frac{a^{\prime\prime}}{a}+ 2 \Hc^2 \gamma_T 
\right)G^s_\boldmathsymbol{k}(\eta,\bar{\eta}) 
= 
\delta\left(\eta-\bar{\eta}\right),
\eeq
with the boundary conditions $\lim_{\eta\to 
\bar{\eta}}G^s_\boldmathsymbol{k}(\eta,\bar{\eta}) = 0$ and $ \lim_{\eta\to 
\bar{\eta}}G^{s,\prime}_\boldmathsymbol{k}(\eta,\bar{\eta})=1$.  

Having extracted above the tensor perturbations, the next step is to derive the 
tensor power spectrum, $\mathcal{P}_{h}(\eta,k)$, for the different 
polarization 
modes, which is defined as the equal-time correlator of the tensor 
perturbations 
through the following relation:
\bea\label{tesnor power spectrum definition}
\langle 
h^r_{\boldmathsymbol{k}}(\eta)h^{s,*}_{\boldmathsymbol{k}^\prime}(\eta)\rangle 
\equiv \delta^{(3)}(\boldmathsymbol{k} - \boldmathsymbol{k}^\prime) \delta^{rs} 
\frac{2\pi^2}{k^3}\mathcal{P}^s_{h}(\eta,k),
\eea
where $s=(\times)$ or $(+)$.
Finally,  after a long but straightforward calculation one acquires 
that $\mathcal{P}_{h}(\eta,k)$ for the $(\times)$ and $(+)$ polarization states 
can be recast as 
~\cite{Ananda:2006af,Baumann:2007zm,Kohri:2018awv,Espinosa:2018eve} 
\bea
\label{Tensor Power Spectrum}
\mathcal{P}^{(\times)\;\mathrm{or}\; (+)}_h(\eta,k) = 4\int_{0}^{\infty} 
\mathrm{d}v\int_{|1-v|}^{1+v}\mathrm{d}u \left[ \frac{4v^2 - 
(1+v^2-u^2)^2}{4uv}\right]^{2}I^2(u,v,x)\mathcal{P}_\Phi(kv)\mathcal{P}
_\Phi(ku)\,.
\eea 
 The two auxiliary variables $u$ and $v$ are defined as $u \equiv 
|\boldmathsymbol{k} - \boldmathsymbol{q}|/k$ and $v \equiv q/k$ and the kernel 
function $I(u,v,x)$ is given by
\bea
\label{I function}
I(u,v,x) = \int_{x_\mathrm{d}}^{x} \mathrm{d}\bar{x}\, 
\frac{a(\bar{x})}{a(x)}\, 
k\, G^s_{k}(x,\bar{x}) F_k(u,v,\bar{x}).
\eea
In the above expressions $x=k\eta$, and we use the notation 
$F_{k}(u,v,\eta)\equiv  F(k ,|\boldmathsymbol{k}-\boldmathsymbol{q}|,\eta)$ 
since  the function $F(\boldmathsymbol{q},\boldmathsymbol{k-q},\eta)$ depends 
only on the modulus of its first two arguments.

\subsection{The gravitational-wave energy-density spectrum}\label{subsec:rho_GW}

In this subsection we calculate   the energy density associated to the 
scalar induced GWs, 
focusing only on subhorizon scales. Consequently, after a 
lengthy but straightforward calculation the GW energy density can be 
recast as \cite{Maggiore:1999vm}
\bea
\label{rho_GW effective}
 \rhoGW (\eta,\boldmathsymbol{x}) =  \frac{\Mp^2}{32 a^2}\, 
\overline{\left(\partial_\eta h_\mathrm{\alpha\beta}\partial_\eta 
h^\mathrm{\alpha\beta} +  \partial_{i} 
h_\mathrm{\alpha\beta}\partial^{i}h^\mathrm{\alpha\beta} \right)}\, ,
\eea
which is simply the sum of a kinetic term and a gradient term.  The overall bar 
denotes an oscillation averaging on sub-horizon scales, performed   to deduce 
the envelope of the gravitational-wave spectrum. The GW spectral 
abundance is just the GW energy density per logarithmic comoving scale, i.e. 
\beq\label{Omega_GW}
\Omega_\mathrm{GW}(\eta,k) = 
\frac{1}{\bar{\rho}_\mathrm{tot}}\frac{\mathrm{d}\rho_\mathrm{GW}(\eta,k)}{
\mathrm{d}\ln k}.
\eeq

Considering a matter-dominated era driven by PBHs, where $w=0$, the transfer 
function $T_\mathrm{\Phi}$ is constant in time (see the discussion after 
\Eq{Bardeen 
potential 2}) , 
and we normalise it to one at PBH domination time, namely 
$T_\mathrm{\Phi}(x_\mathrm{d})=1$. This forces the source term 
$S^s_\boldmathsymbol{k}$ to be constant in time and consequently at 
sub-horizon scales, where $k\gg \cal{H}$, from \Eq{Tensor Eq. of Motion} we 
obtain that $h^s_\boldmathsymbol{k}\simeq 
\frac{4S^s_\boldmathsymbol{k}}{k^2}$. 
Finally, the tensor modes have a mild dependence on time and therefore the 
kinetic term in the expression for the GW energy density (\ref{rho_GW 
effective}) can be neglected. Therefore, we straightforwardly obtain 
that 
\beq\label{rho_GW_effective MD}
\begin{split}
\left\langle \rhoGW (\eta,\boldmathsymbol{x}) \right\rangle &  \simeq 
\left\langle \rho_{\mathrm{GW,grad}} (\eta,\boldmathsymbol{x}) \right\rangle = 
\sum_{s=+,\times}\frac{\Mp^2}{32a^2}\overline{
\left\langle\left(\nabla h^{s}_\mathrm{\alpha\beta}\right)^2\right \rangle }
   \\ & =   \frac{\Mp^2}{32a^2 \left(2\pi\right)^3} 
\sum_{s=+,\times} \int\mathrm{d}^3\boldmathsymbol{k}_1 
\int\mathrm{d}^3\boldmathsymbol{k}_2\,  k_1 k_2 \overline{  \left\langle 
h^{s}_{\boldmathsymbol{k}_1}(\eta)h^{s,*}_{\boldmathsymbol{k}_2}
(\eta)\right\rangle} e^{i(\boldmathsymbol{k}_1-\boldmathsymbol{k}_2)\cdot 
\boldmathsymbol{x}}\,,
 \end{split}
\eeq
where the brackets stand for an ensemble average.
At the end, by combining \Eq{rho_GW_effective MD}, \Eq{Omega_GW} and \Eq{tesnor 
power spectrum definition} and taking into account from \Eq{Tensor Power 
Spectrum} that the $(\times)$ and $(+)$ polarization modes give an equal 
contribution, we find that 
\beq\label{Omega_GW_sub_horizon}
\Omega_\mathrm{GW}(\eta,k) \simeq  
\frac{1}{\bar{\rho}_\mathrm{tot}}\frac{\mathrm{d}\rho_\mathrm{GW,grad}(\eta,k)}{
\mathrm{d}\ln k} =  
\frac{1}{48}\left(\frac{k}{\calH(\eta)}\right)^{2}\overline{\mathcal{P}^{
(\times)}_h(\eta,k)}.
\eeq

In order to compute the contribution of the induced GWs to the energy 
budget of the Universe at the present epoch, one should evolve 
$\Omega_\mathrm{GW}(\eta,k)$ from a reference conformal time $\eta_\mathrm{*}$ 
up to today. To do so, one has that 
\beq
\OmegaGW(\eta_0,k) = \frac{\rhoGW(\eta_0,k)}{\rho_\mathrm{c}(\eta_0)} = 
\frac{\rhoGW(\eta_\mathrm{*},k)}{\rho_\mathrm{c}(\eta_\mathrm{*})}\left(\frac{
a_\mathrm{*}}{a_\mathrm{0}}\right)^4 
\frac{\rho_\mathrm{c}(\eta_\mathrm{*})}{\rho_\mathrm{c}(\eta_0)} = 
\OmegaGW(\eta_\mathrm{*},k)\Omega^{(0)}_\mathrm{r}\frac{\rho_\mathrm{r,*}
a^4_\mathrm{*}}{\rho_\mathrm{r,0}a^4_0},
\eeq
where we have taken into account that $\Omega_\mathrm{GW}\sim a^{-4}$, and 
where the  
index 
$0$ refers to the present time. Then, taking into account that the energy 
density of radiation can be recast as $\rho_r = 
\frac{\pi^2}{15}g_{*\mathrm{\rho}}T_\mathrm{r}^4$ and that the temperature of 
the radiation bath, $T_\mathrm{r}$, scales as $T_\mathrm{r}\propto 
g^{-1/3}_{*\mathrm{S}}a^{-1}$, one finds that 
\beq\label{Omega_GW_RD_0}
\Omega_\mathrm{GW}(\eta_0,k) = 
\Omega^{(0)}_r\frac{g_{*\mathrm{\rho},\mathrm{*}}}{g_{*\mathrm{\rho},0}}
\left(\frac{g_{*\mathrm{S},\mathrm{0}}}{g_{*\mathrm{S},\mathrm{*}}}\right)^{4/3}
\OmegaGW(\eta_\mathrm{*},k),
\eeq
where $g_{*\mathrm{\rho}}$ and $g_{*\mathrm{S}}$ stand for the energy and 
entropy relativistic degrees of freedom.

\section{Constraints on $f(T)$ gravity}
\label{ConstrfT}

In this section  we use the portal of the scalar induced GWs from PBH Poisson 
fluctuations presented  above, in order to derive constraints on $f(T)$  
gravity.

\subsection{Mono-parametric $f(T)$ Models}\label{subsec:f(T) models}

Since we will perform specific 
calculations we have to specify the form of the function $f(T)$. In particular, 
we consider the following three $f(T)$ gravity models depending on one free
parameter, denoted here as $\beta$.
\begin{enumerate}
\item The power-law model \cite{Bengochea:2008gz}
(hereafter $f_{1}$ model), in which 
\begin{equation}
f(T)=T+\alpha (-T)^{\beta},
\label{powermod}
\end{equation} 
with
\begin{eqnarray}
\alpha=(6H_0^2)^{1-\beta}\frac{\Omega_{F0}}{2\beta-1},
\end{eqnarray}
where $\Omega_{F0}=1-\Omega_{m0}-\Omega_{r0}$. According to  
observational constraints for $\beta$ one has that 
$-0.3<\beta<0.3$ \cite{Nesseris:2013jea,Nunes:2018evm,Anagnostopoulos:2019miu}. 
Note that  GR is recovered for 
$\beta\rightarrow0$.

\item The square-root exponential model (hereafter $f_{2}$) 
\cite{Linder:2010py}
\begin{eqnarray}
f(T)=T+\alpha T_{0}(1-e^{-\frac{1}{\beta}\sqrt{T/T_{0}}}),
\label{Lindermod}
\end{eqnarray}
with 
\begin{eqnarray}
\alpha=\frac{\Omega_{F0}}{1-(1+\frac{1}{\beta})e^{-\frac{1}{\beta}}}.
\end{eqnarray} 
The $\beta$ parameter is observationally constrained within the range 
$0.05<\beta<0.4$ \cite{Nesseris:2013jea,Nunes:2018evm,Anagnostopoulos:2019miu}  
and GR is recovered for 
$\beta\rightarrow0^+$.

\item  The   exponential model (hereafter $f_{3}$) \cite{Nesseris:2013jea}:
\begin{eqnarray}
f(T)=T+\alpha T_{0}[1-e^{-T/(\beta T_{0})}],
\label{f3cdmmodel}
\end{eqnarray}
with
\begin{eqnarray}
\alpha=\frac{\Omega_{F0}}{1-(1+\frac{2}{\beta})e^{-\frac{1}{\beta}}}.
\end{eqnarray}
The $\beta$ parameter is observationally constrained within the range 
$0.02<\beta<0.2$  \cite{Nesseris:2013jea,Nunes:2018evm,Anagnostopoulos:2019miu} 
and GR is recovered for 
$\beta\rightarrow0^+$.
\end{enumerate}
 
\subsection{Theoretical parameters }

In this subsection we discuss  the
theoretical parameters that are involved in the analysis. These parameters 
include the mass of the 
PBH $m_\mathrm{PBH}$, the initial PBH abundance at 
formation time $\Omega_\mathrm{PBH,f}$, and of course the parameter $\beta$ of 
the 
mono-parametric $f(T)$ model at hand.

For the PBH mass range we consider that the PBHs     are formed 
after the end of inflationary era and evaporate before the BBN time. In 
particular, one can
derive     an upper bound on the PBH mass   by 
accounting for the current Planck upper bound on the tensor-to-scalar ratio for 
single-field slow-roll models of inflation, which gives  
$\rho^{1/4}_\mathrm{inf}<10^{16}\mathrm{GeV}$ ~\cite{Akrami:2018odb}. On the 
other hand one can extract a conservative lower bound on the reheating energy 
scale, i.e. 
$\rho^{1/4}_\mathrm{reh}> 4\mathrm{MeV}$,  derived by taking into account the 
thermalization of neutrino background and the hadron scatterings emitted from 
PBHs as discussed in~\cite{Hasegawa:2019jsa}. Hence, one can show that 
   the relevant PBH mass range is given by ~\cite{Papanikolaou:2020qtd}
\bea
\label{eq:domain:mPBHf}
10 \mathrm{g}< m_\mathrm{PBH}< 10^{9} \mathrm{g}.
\eea
 
We proceed to  the range of $\Omega_\mathrm{PBH,f}$. In order to 
have a 
transient PBH domination era, we can set it by requiring that the PBH 
evaporation time 
$t_\mathrm{evap}$  is larger than the PBH domination time $t_\mathrm{d}$. 
Consequently, knowing that during a radiation dominated era we have
$\Omega_\mathrm{PBH}=\rho_\mathrm{PBH}/\rho_\mathrm{d} \propto 
a^{-3}/a^{-4}\propto a$ and demanding  that 
$t_\mathrm{evap}>t_\mathrm{d}$, we obtain that 
\bea
\label{eq:domain:OmegaPBHf}
\Omega_\mathrm{PBH,f} > 10^{-15} \sqrt{\frac{g_\ueff}{100}} 
\frac{10^9\mathrm{g}}{m_\mathrm{PBH}}\, .
\eea

Lastly, concerning  the   parameter $\beta$, according to 
observational constraints mentioned above from  
\cite{Nesseris:2013jea,Nunes:2018evm,Anagnostopoulos:2019miu}, its value 
should 
roughly vary within the following range:
\bea
\label{eq:domain:alpha}
-0.4 \leq \beta\leq 0.4\, ,
\eea
depending on the $f(T)$ model at hand.

\subsection{Gravitational waves from an early primordial black hole dominated 
era}

We have now all the necessary material in order to  investigate   the 
relevant GW signal created in the  early PBH dominated 
era, in the context of 
$f(T)$ gravity. Let us mention here that one should discriminate between two 
decisive effects 
which introduce deviations from standard GR gravity, namely a) the effect of 
the source of the induced GWs, which is actually encapsulated within the power 
spectrum of the PBH gravitational potential $\mathcal{P}_\Phi$, and b) the 
effect 
of the GW propagation, which is encoded within the time evolution of the Green 
function $G_k(\eta,\bar{\eta})$, which can be viewed as the propagator of the 
tensor perturbations as it can be seen by \Eq{tensor mode function}.

\subsubsection{The effect of    the gravitational-wave   source}

After solving numerically the Meszaros equation  (\ref{eq:Meszaros in 
f(T):subhorizon}) for the PBH energy density perturbations, we proceed to the 
calculation of the 
PBH gravitational potential power spectrum $\mathcal{P}_\Phi$, as dictated by 
\Eq{eq:PowerSpectrum:Phi:PBHdom:f(T)}, by accounting for the three 
mono-parametric $f(T)$ models presented in \Sec{subsec:f(T) models}. The power 
spectrum $\mathcal{P}_\Phi$ is actually the source of the induced GWs as it can 
be seen from \Eq{Tensor Power Spectrum}. 


\begin{figure}\centering
\subfloat[The power-law 
model]{\label{a}\includegraphics[width=.45\linewidth]{
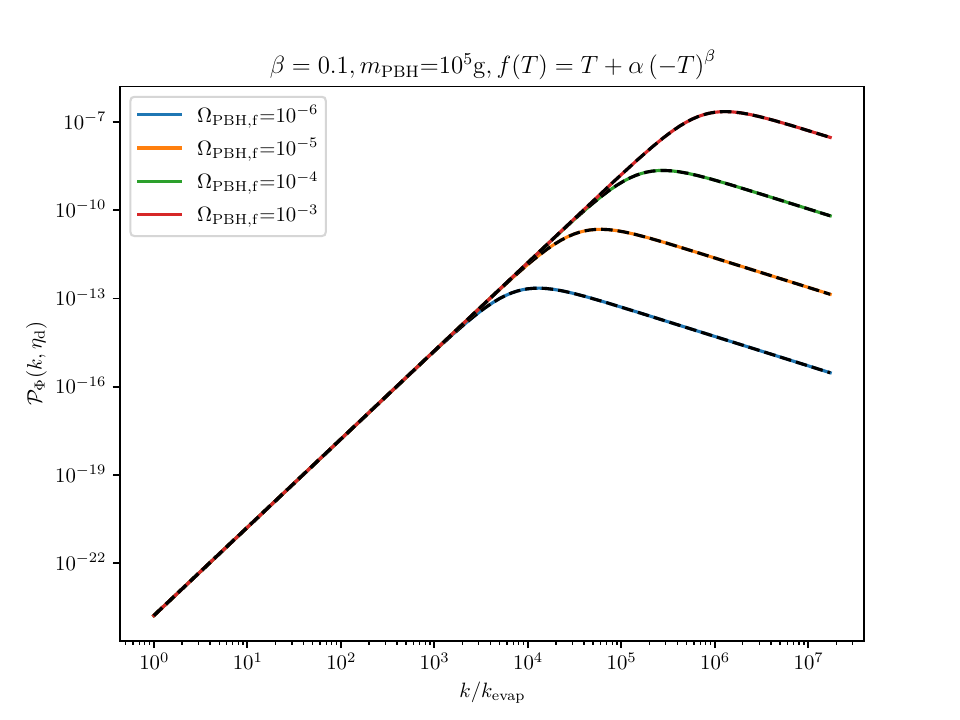}}\hfill
\subfloat[The square-root exponential
model]{\label{b}\includegraphics[width=.45\linewidth]{
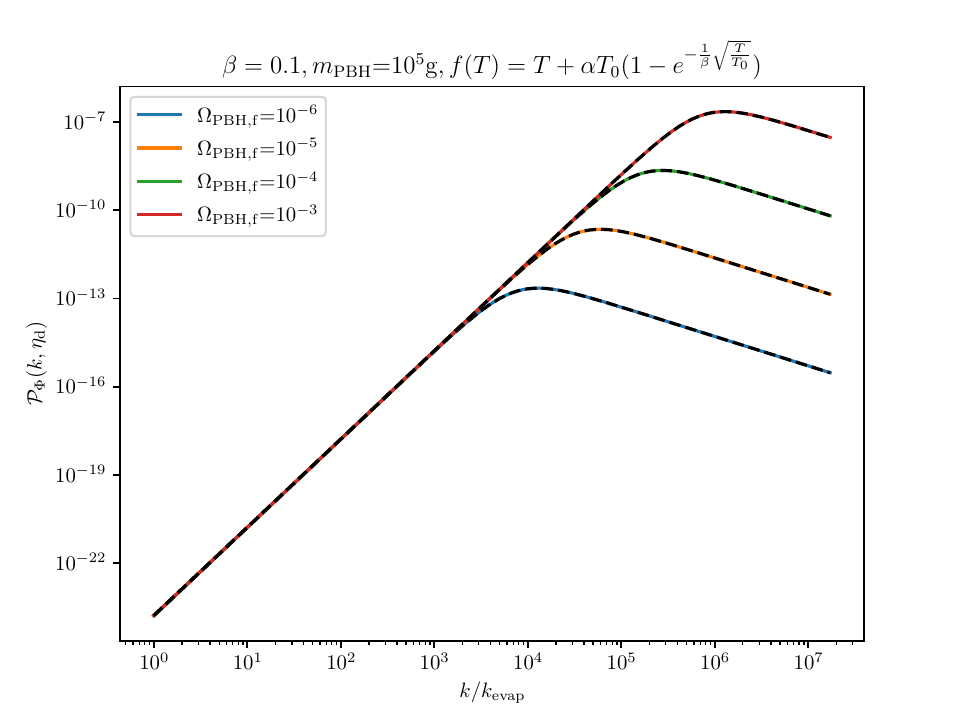}}\par 
\subfloat[The exponential 
model]{\label{c}\includegraphics[width=.45\linewidth]{
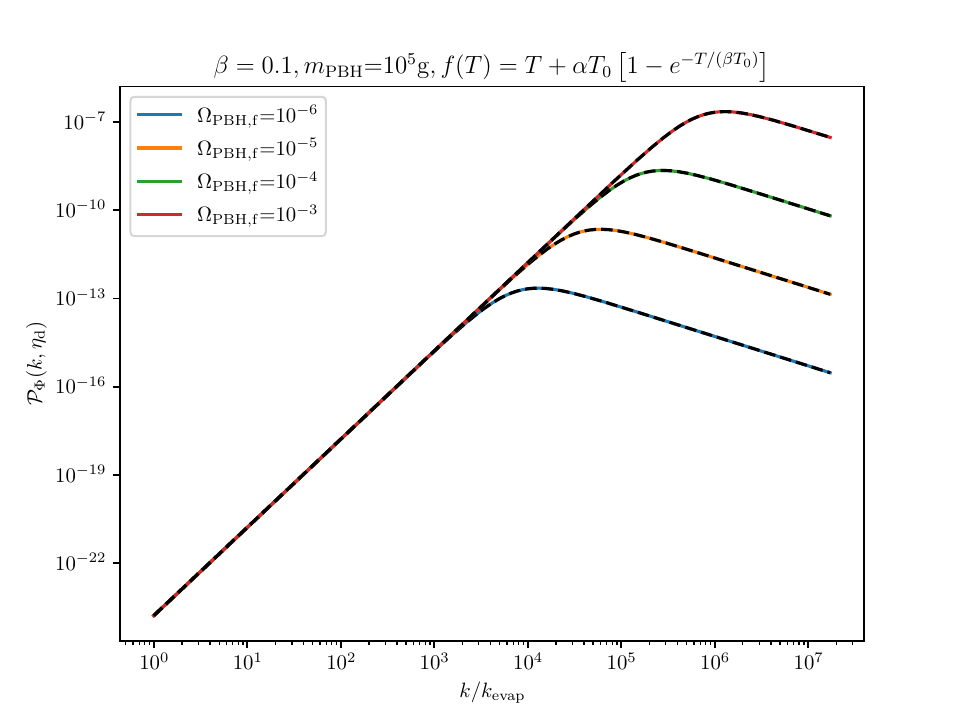}}
\caption{{\it{The power spectrum of the PBH 
gravitational 
potential at PBH   domination time, for the three mono-parametric 
$f(T)$ 
models, by choosing different values of the initial PBH 
abundance 
$\Omega_\mathrm{PBH,f}$. In all   graphs  we have used $\beta=0.1$ and 
$m_\mathrm{PBH,f}=10^{5}\mathrm{g}$. The dashed curves correspond to the GR 
results.}}}
\label{fig:P_Phi_OmegaPBHf}
\end{figure}

\begin{figure}\centering
\subfloat[The power-law 
model]{\label{a}\includegraphics[width=.45\linewidth]{
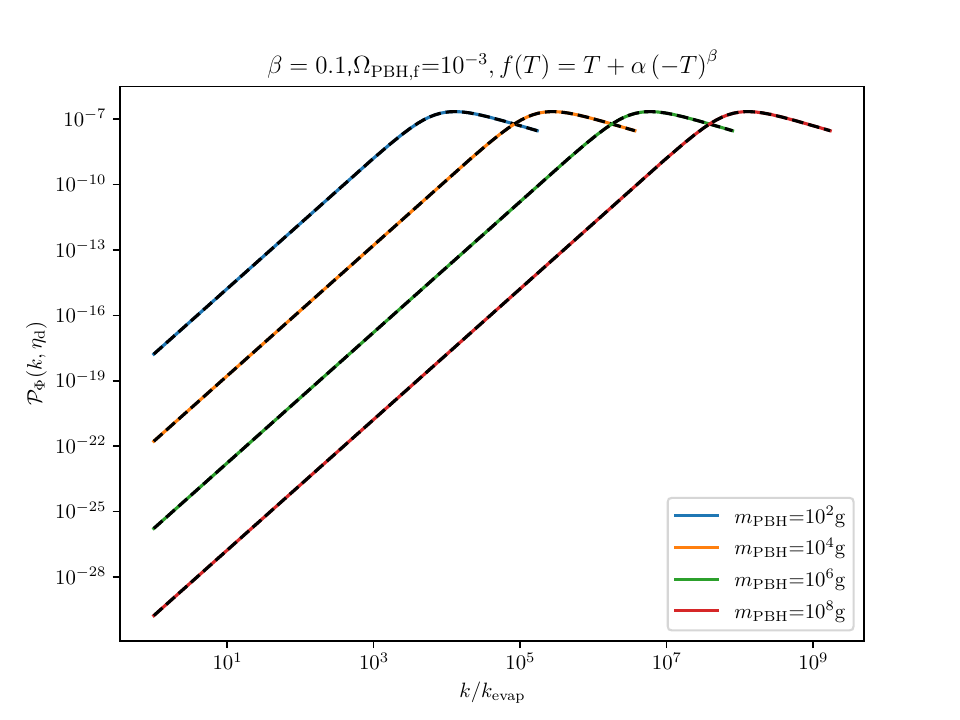}}\hfill
\subfloat[The square-root exponential 
model]{\label{b}\includegraphics[width=.45\linewidth]{
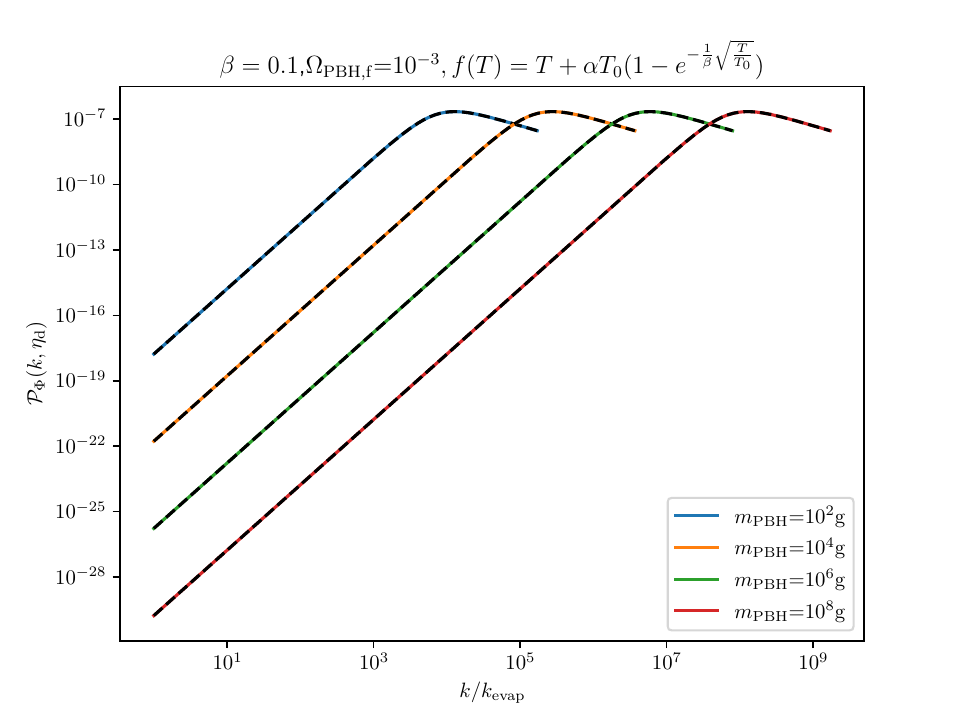}}\par 
\subfloat[The exponential 
model]{\label{c}\includegraphics[width=.45\linewidth]{
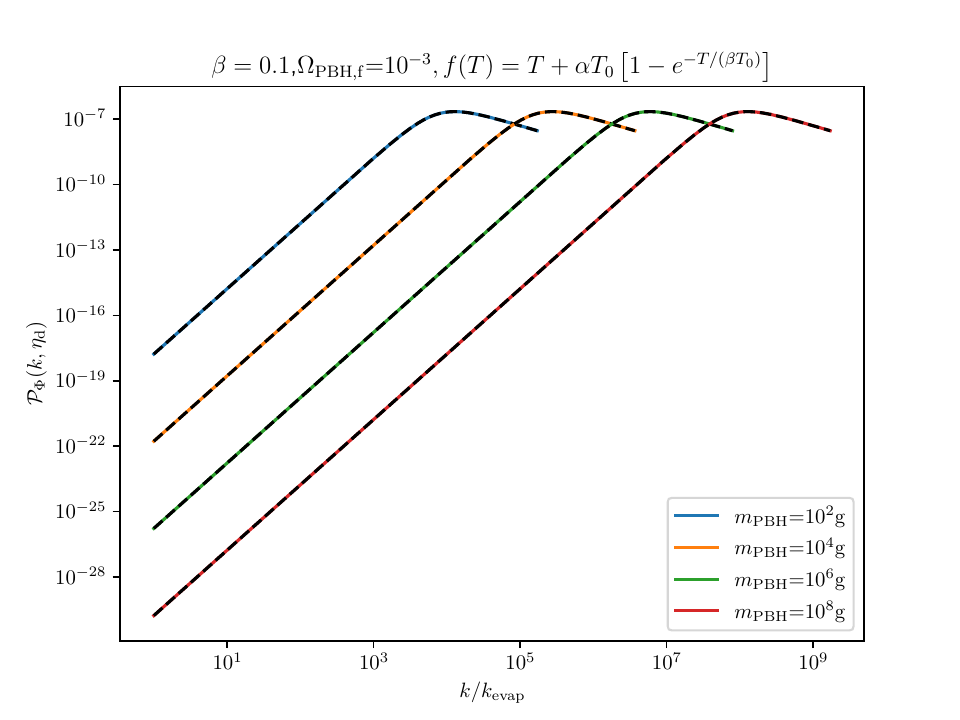}}
\caption{{\it{The power spectrum of the PBH 
gravitational 
potential at PBH   domination time, for the three mono-parametric 
$f(T)$ 
models,  by choosing different values of the  PBH mass 
$m_\mathrm{PBH}$. In all   graphs  we have used $\beta=0.1$ and 
$\Omega_\mathrm{PBH,f}=10^{-3}$. The dashed curves correspond to the GR 
results.}}}
\label{fig:P_Phi_mPBH}
\end{figure}

\begin{figure}\centering
\subfloat[The power-law 
model]{\label{a}\includegraphics[width=.45\linewidth]{
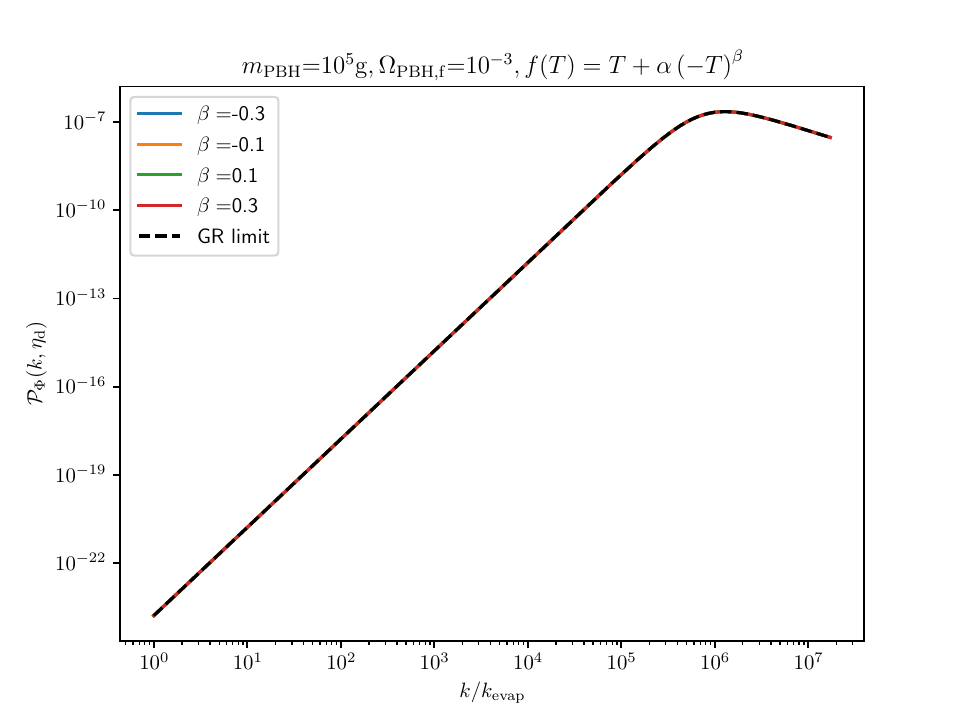}}\hfill
\subfloat[The square-root exponential 
model]{\label{b}\includegraphics[width=.45\linewidth]{
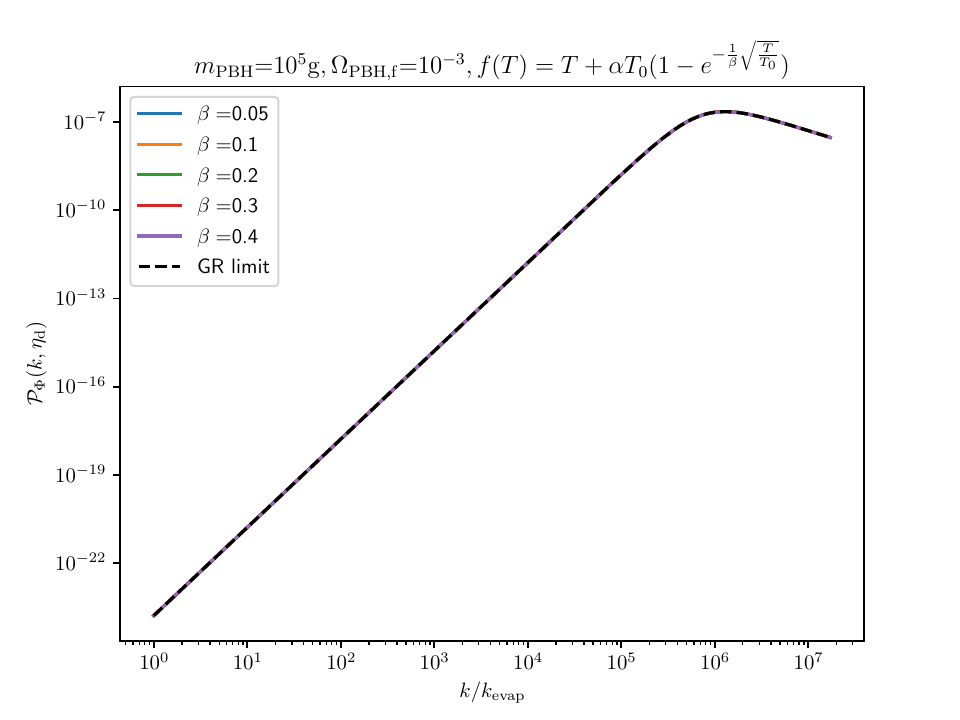}}\par 
\subfloat[The exponential 
model]{\label{c}\includegraphics[width=.45\linewidth]{
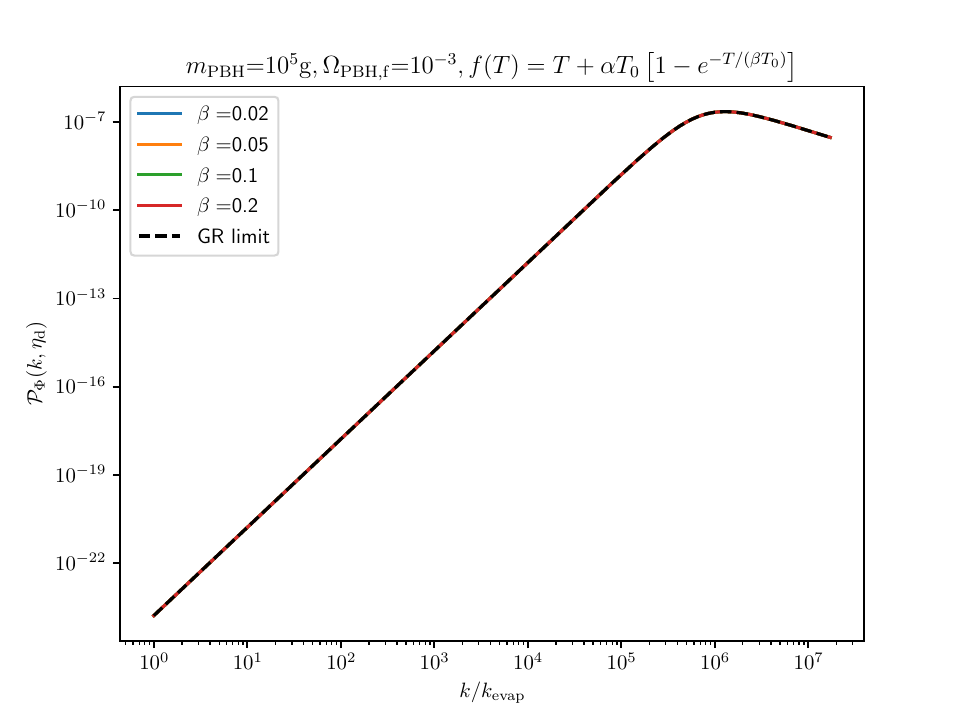}}
\caption{{\it{The power spectrum of the PBH 
gravitational 
potential at PBH   domination time, for the three mono-parametric 
$f(T)$ 
models,  by choosing different values of the parameter $\beta$ 
within its observationally allowed range. In all   graphs  we have used 
$m_\mathrm{PBH}=10^5\mathrm{g}$ and $\Omega_\mathrm{PBH,f}=10^{-3}$. The dashed 
curves correspond to the GR results.}}}
\label{fig:P_Phi_mPBH_OmegaPBHf}
\end{figure}

In Figs. \ref{fig:P_Phi_OmegaPBHf}-\ref{fig:P_Phi_mPBH_OmegaPBHf} we depict 
the power spectrum of the PBH gravitational 
potential at PBH   domination time, for the three  
$f(T)$ models,  for various choices  of the initial PBH 
abundance 
$\Omega_\mathrm{PBH,f}$, of the  PBH mass 
$m_\mathrm{PBH}$, and of the model-parameter $\beta$.
As it can be clearly seen from \Fig{fig:P_Phi_OmegaPBHf}, the amplitude of the 
PBH gravitational potential power spectrum increases with the initial PBH 
abundance, whereas from  \Fig{fig:P_Phi_mPBH} we can notice that the position 
of 
the peak of $\mathcal{P}_\Phi$ depends on the PBH mass. Finally, from 
\Fig{fig:P_Phi_mPBH_OmegaPBHf}, where $\mathcal{P}_\Phi$ is plotted for 
different values of the modified gravity parameter $\beta$, one can clearly 
infer that the deviation from GR is practically indistinguishable. As it was 
verified numerically for all the monoparametric $f(T)$ models considered here, the relative change of the gravitational potential power 
spectrum with respect to the one of GR around the peak of the power spectrum, 
i.e. at $k\sim k_\mathrm{d}$, is of the order $10^{-2}$, namely
\beq
\left\vert\frac{\mathcal{P}^{f(T)}_\Phi(k_\mathrm{d}) 
-\mathcal{P}^{GR}_\Phi(k_\mathrm{d}) 
}{\mathcal{P}^{GR}_\Phi(k_\mathrm{d})}\right\vert\sim 10^{-2}.
\eeq

In summary, we conclude that the effect of $f(T)$ gravity on the source of 
gravitational waves, is very small, and practically indistinguishable from GR 
for realistic $f(T)$ model-parameter values.

\subsubsection{The effect of the gravitational-wave propagation}

We now come to the effect of the GW propagation. One should 
extract the 
behavior 
of $G_k(\eta,\bar{\eta})$ by solving \Eq{eq:Green function equation in f(T)} 
and 
investigate possible deviations from GR. During the time evolution of the GW 
spectrum one should take into account the fact that during a sudden transition 
from the PBH dominated era to the subsequent radiation era, in the case of a 
monochromatic PBH mass function as the one considered here, the GW is enhanced 
due a very rapid increase of the time derivative of the gravitational potential 
$\Phi$ which is present in the source term (see \Eq{Source}) as noted in 
~\cite{Inomata:2019ivs,Domenech:2021wkk}. Then, during the radiation-dominated 
era, the source term is decaying on subhorizon scales and therefore 
$\Omega_\mathrm{GW}$ stops growing after the moment when the source term has 
sufficiently decayed. After this point, the scalar induced GWs are propagating 
as free waves, with their present energy density spectrum given by 
\Eq{Omega_GW_RD_0} and $\eta_{*}$ being a reference time during the 
radiation-dominated era when GWs start to propagate as free waves.

Having these in mind let us focus on \Eq{eq:Green function equation in f(T)}, 
namely 
\beq\label{eq:Green_function_propagation_effect}
G_\boldmathsymbol{k}^{s,\prime\prime}(\eta,\bar{\eta})  - 2\Hc \gamma_T 
G_\boldmathsymbol{k}^{s,\prime}(\eta,\bar{\eta}) + \left( k^{2}  
-\frac{a^{\prime\prime}}{a}+ 2 \Hc^2 \gamma_T 
\right)G^s_\boldmathsymbol{k}(\eta,\bar{\eta}) 
= 
\delta\left(\eta-\bar{\eta}\right).
\eeq
One   can identify the dominant terms of the above equation by taking the 
ratios between 
the GR terms and the new $f(T)$
terms multiplied by the 
$\gamma_T$ function. In 
this procedure, we should take into consideration the fact that the $\gamma_T$ 
function, as it was checked numerically for all the three mono-parametric 
models 
studied here, is a negative decreasing function of time, which implies that its 
absolute value increases with time. Therefore, in order to find the maximum 
deviation from GR we compute the ratios between the GR and $f(T)$ terms at a 
time during radiation domination when the $\gamma_T$ function acquires its 
maximum value. Being 
quite 
conservative we set this time to be the standard matter-radiation equality 
time at redshift $z_\mathrm{eq}=3387$.

At this point, we should stress   that in the comparison of the different 
terms in \Eq{eq:Green_function_propagation_effect} one should compute the 
derivative terms $G^\prime_k(\eta,\bar{\eta})$ and 
$G^{\prime\prime}_k(\eta,\bar{\eta})$. To achieve this we take into account the 
fact 
that the solutions of \Eq{eq:Green_function_propagation_effect} are expected to 
be trigonometric functions (sines and cosines), in particular Bessel functions 
in the case of GR. One then expects that $G^\prime_k(\eta,\bar{\eta})$, 
$G_k(\eta,\bar{\eta})$ and $G^{\prime\prime}_k(\eta,\bar{\eta})$ should differ 
by a phase difference, and this was indeed verified numerically. Consequently, 
one expects in general that $\left\vert 
G^\prime_k(\eta,\bar{\eta})/G_k(\eta,\bar{\eta})\right\vert_{\eta=\eta_\mathrm{
eq}}\sim O(1)$ and $\left\vert 
G^{\prime\prime}_k(\eta,\bar{\eta})/G_k(\eta,\bar{\eta})\right\vert_{
\eta=\eta_\mathrm{eq}}\sim O(1)$,  when comparing the different terms of 
\Eq{eq:Green_function_propagation_effect}.

Thus, taking the above discussion into account, let us start   the 
identification of the dominant terms by comparing   the first two terms 
in \Eq{eq:Green_function_propagation_effect}, namely the second derivative term 
$G_\boldmathsymbol{k}^{s,\prime\prime}(\eta,\bar{\eta})$ and the friction term 
$2\Hc \gamma_T G_\boldmathsymbol{k}^{s,\prime}(\eta,\bar{\eta})$, and in 
particular let us  consider 
their 
ratio $G_\boldmathsymbol{k}^{s,\prime\prime}(\eta,\bar{\eta})/[2\Hc \gamma_T 
G_\boldmathsymbol{k}^{s,\prime}(\eta,\bar{\eta})]$. Eventually,  we find 
that 
for the power-law $f(T)$ model and for $m_\mathrm{PBH}=10^{5}\mathrm{g}$, 
$\Omega_\mathrm{PBH,f}=10^{-3}$ and $\beta=0.1$, we acquire 
\beq
\left\vert\frac{G_\boldmathsymbol{k}^{\prime\prime}(\eta,\bar{\eta})}{2\Hc 
\gamma_T G_\boldmathsymbol{k}^{\prime}(\eta,\bar{\eta})}\right\vert \simeq 
\left.\frac{1}{2\Hc \gamma_T}\right\vert_{\eta=\eta_\mathrm{eq}} \simeq 
10^{46}\gg 1.
\eeq
Similar results are obtained for the square-root exponential and the 
exponential 
$f(T)$ 
models, and by varying  the parameters $m_\mathrm{PBH}$, 
$\Omega_\mathrm{PBH,f}$ and $\beta$ too.

With the same reasoning we can additionally examine  the ratio between
the $k^2$ and 
$2\mathcal{H}^2\gamma_T$ terms inside the parenthesis of 
\Eq{eq:Green_function_propagation_effect}. Choosing again 
$\eta=\eta_\mathrm{eq}$ in order to find the maximum deviation from GR,  we 
straightforwardly find that  
\beq
\left. \frac{k^2}{2\Hc^2 \gamma_T 
}\right\vert_{k=k_\mathrm{evap},\eta=\eta_\mathrm{eq}} \simeq 10^{83}\gg 1,
\eeq
where $k=k_\mathrm{evap}$ is the comoving scale exiting the Hubble radius at 
PBH 
evaporation time and as a consequence it is the largest scale considered here. 

In summary, we  can safely argue that the $f(T)$ modifications at the level of 
the 
propagation equation \eqref{eq:Green_function_propagation_effect} can be 
neglected and consequently one obtains that
\beq
G_\boldmathsymbol{k}^{f(T)}
(\eta,\bar{\eta})\simeq
G_\boldmathsymbol{k}^{\mathrm{GR}}
(\eta,
{\bar{\eta}}).
\eeq
Hence,  we conclude that the effect of $f(T)$ gravity on the propagation of 
gravitational waves, is very small, and practically indistinguishable from GR 
for realistic $f(T)$ model-parameter values. We mention that this behavior is different than the case of $f(R)$ gravity, in which the corresponding effect is small but still distinguishable from GR \cite{Papanikolaou:2021uhe}, which reveals the different nature and effects of the two gravitational modifications.

\section{Conclusions}
\label{Conclusions}

Primordial black holes (PBH) can address a number of issues of modern cosmology 
since they may account 
for a part or all of the dark matter contribution, they can seed the 
large-scale 
structure formation through Poisson fluctuations, and they may constitute   the 
progenitors of the black-hole merging events recently detected by LIGO-VIRGO. 
Interestingly, PBHs are tightly associated with gravitational wave (GW) 
signals, providing  us the possibility to gain information of the physics of 
different cosmic epochs,
from the very early universe up to later times, depending on the GW production 
mechanism. Since their formation and effects are determined by the underlying 
gravitational theory, one can use them as a novel tool 
in order to test general 
relativity and investigate possible modified gravity deviations.

In this work we focused on the primordial scalar induced gravitational waves, 
generated at second order in cosmological 
perturbation theory, from PBH Poisson 
fluctuations, in the framework of $f(T)$ modified gravity. In particular, we 
desired to use it as a novel probe to extract constraints on the involved 
model-parameters.  We  considered three viable mono-parametric $f(T)$ models, 
and we investigated the induced modifications at the level of 
the gravitational-wave source, which are encoded in terms of the power 
spectrum of the PBH gravitational potential $\mathcal{P}_\Phi$, as well as at 
the level of their propagation, described in terms 
of the Green function $G_\boldmathsymbol{k}(\eta,\bar{\eta})$ which can be 
considered as the propagator of the tensor perturbations.

Our detailed analysis showed that  within the observationally allowed range of 
the parameters of the $f(T)$ models at hand, the obtained deviations from GR,
both at the level of source, as well as at the level of propagation,
are  practically indistinguishable. Indicatively, regarding the PBH gravitational 
potential power spectrum we found that the deviation from GR is of the order of 
$10^{-2}$ for all the monoparametric $f(T)$ models considered here. This behavior is different than the case of other modified gravity theories, such as $f(R)$ gravity, in which the corresponding effect is small but still distinguishable from GR \cite{Papanikolaou:2021uhe}. Hence, we conclude that realistic and viable $f(T)$ theories of 
gravity can safely pass the primordial black hole constraints, which may offer 
an additional argument in their favor.

Finally, one should stress that one can extend our analysis to other modified teleparallel theories of gravity, such as $f(T,B)$ gravity  and scalar-torsion theories, whose polarization  numbers have been calculated in \cite{Abedi:2017jqx}. Especially in the cases where extra polarization modes do appear, one expects to find significant differences, as it was found in \cite{Papanikolaou:2021uhe}. This interesting and necessary investigation will be performed in a separate paper.

\begin{acknowledgments}
T.P. acknowledges financial support from the Foundation for Education
and European Culture in Greece. The authors would like to acknowledge the 
contribution of the COST Action 
CA18108 ``Quantum Gravity Phenomenology in the multi-messenger approach''.

\end{acknowledgments}

\appendix

 \section{Super-horizon scales in $f(T)$ gravity} 
 \label{AppendixB}
 
 In this Appendix we examine the behavior of pertrubations at 
super-horizon scales in the framework of $f(T)$ gravity. We   use the 
  definition of comoving curvature perturbation as
 \begin{equation}
     \mathcal{R} \equiv - \Phi - \mathcal{H} \upsilon \label{R}.
 \end{equation} 
 At super-Hubble scales, \Eq{delta_00_fT_bounce} under the assumptions 
$F_T\ll1$ 
and $F_{TT}\ll 1$ becomes:
 \beq
 3\mathcal{H}(\Phi' + \mathcal{H}\Psi) 
= -4\pi G a^2 \, \delta \rho,
\eeq
and thus together with \Eq{delta_0i_fT_bounce} 
yields:
  \begin{equation}
     \mathcal{R} = - \Phi + \frac{\delta}{3(1+w)} 
\xrightarrow{(\ref{zeta})} - \zeta, \quad k\ll \mathcal{H}. \label{RZ}
 \end{equation}
Furthermore, from \Eq{R} and \Eq{delta_0i_fT_bounce} we can write:
\begin{equation}
    \mathcal{R}  = \Phi + \frac{\mathcal{H}({\Phi}' + \mathcal{H} \Psi)}{4\pi G 
a^2 \bar{\rho} (1+w)} \xrightarrow{\mathcal{H}^2 = 8 
\pi G a^2 \bar{\rho}/3}  \Phi + \frac{2}{3} \frac{\Phi'/\mathcal{H} 
+ \Psi}{ 1+w} \label{Raf}.
\end{equation}
Moreover, from (\ref{delta_ij_fT_bounce})  by 
neglecting the anisotropic stress we see that $ \Phi = \Psi $ . 
Therefore, under the aforementioned 
assumptions and for 
$k\ll \mathcal{H}$, we  obtain that at super-horizon scales we have
\bea
\mathcal{R}  = 
\frac{2}{3}\frac{{\Phi}^\prime/\mathcal{H}+\Phi}{1+w}+\Phi\, .
\eea

\bibliographystyle{JHEP} 
\bibliography{PBH}

\end{document}